\documentclass[3p,times]{elsarticle}

\usepackage{amssymb}
\usepackage{amsmath}
\usepackage{times}
\usepackage{setspace}
\usepackage{geometry}
\usepackage{enumitem}
\usepackage{hyperref}
\usepackage{algorithm}
\usepackage{algpseudocode}
\usepackage{booktabs}
\usepackage{multirow}
\usepackage{graphicx}
\usepackage{subcaption}
\usepackage{amsthm}  % For theorem-like environments
\usepackage{thmtools} % For advanced theorem styling (optional but recommended)
\usepackage{url}
\usepackage{xcolor}

\theoremstyle{definition}  % Use definition style (upright text, not italic)
\newtheorem{observation}{Observation}[section]  % Numbered by section

\journal{Integration}

\begin{document}
	
	\thispagestyle{empty}
	
	\begin{center}
		
		\vspace*{2cm}
		
		{\Large\bfseries Accepted Author Manuscript}
		
		\vspace{1.5cm}
		
		{\LARGE\bfseries
			ExpanderGraph-128: A novel graph-theoretic block cipher with formal security analysis and hardware implementation
		}
		
		\vspace{1.5cm}
		
		{\large
			W. A. S. Wijesinghe
		}
		
		\vspace{2cm}
		
		\begin{spacing}{1.2}
			
			\textbf{Citation:}
			
			W. A. S. Wijesinghe,  
			``ExpanderGraph-128: A novel graph-theoretic block cipher with formal security analysis and hardware implementation,''  
			\textit{Integration, the VLSI Journal}, Elsevier, Volume 109, 	2026, 102715, ISSN 0167-9260, 	https://doi.org/10.1016/j.vlsi.2026.102715.
			
			\vspace{0.7cm}
			
			\textbf{DOI:}
			
			https://doi.org/10.1016/j.vlsi.2026.102715
			
			\vspace{0.7cm}
			
			\textbf{Temporary Free Access (valid until 1 May 2026):}
			
			\url{https://authors.elsevier.com/a/1mljecBfJ8Da\~ }
			
			\vspace{1.5cm}
			
			\textbf{Note:}
			
			This document is the \textbf{Accepted Author Manuscript (AAM)} of the above article.  
			It has been peer reviewed and accepted for publication but has not undergone the
			publisher’s final typesetting, copyediting, or formatting.  
			
			The final published version of record is available via the DOI link above.
			
		\end{spacing}
		
		\vfill
		
	\end{center}

\clearpage

	\begin{frontmatter}
		
		
		
		\title{ExpanderGraph-128: A Novel Graph-Theoretic Block Cipher with Formal Security Analysis and Hardware Implementation}
		
		
		\author{W.A. Susantha  Wijesinghe} %% Author name
		
		\affiliation{organization={Department of Electronics, Wayamba University of Sri Lanka},%Department and Organization
			city={Kuliyapitiya},
			postcode={60200}, 
			country={Sri Lanka}}
		
		%% Abstract

\begin{abstract}
Lightweight block cipher design has converged toward incremental optimization of established paradigms--substitution-permutation networks, Feistel structures, and ARX constructions--where security derives from algebraic complexity of individual components. We introduce a fundamentally different approach: expander-graph interaction networks, where cryptographic security emerges from sparse structural connectivity rather than component sophistication. Unlike Goldreich-type one-way function constructions that use expander graphs as a single-pass hardness mechanism, our design exploits the expander graph as a diffusion primitive within an iterated, keyed confusion--diffusion framework validated through concrete cryptanalysis.

We present ExpanderGraph-128 (EGC128), a 128-bit block cipher instantiating this paradigm through a 20-round balanced Feistel network. Each round applies a 64-bit nonlinear transformation governed by a 3-regular expander graph, where vertices execute identical 4-input Boolean functions on local neighborhoods. Security analysis combines rigorous MILP-based differential bounds, proven optimal via SCIP through 10~rounds, with minimum active Rule-A counts \{4, 13, 29, 53, 85, 125, 173, 229, 291, 355\} establishing 147.3~bits of provable differential security, conservatively extrapolating to 413~bits for the full 20-round cipher, with formal MILP linear trail bounds ($\geq 2^{145}$ distinguisher complexity by conservative extrapolation), related-key analysis proving no free rounds exist for any nonzero key difference, structural attack resistance verified through algebraic degree saturation by round~4 of $F_\mathrm{core}$ and 1{,}800 affine subspace tests finding no invariant structures, and SMT verification. NIST SP~800-22 confirms pseudorandom output quality across $10^8$~bits, with all 15 test categories passing at recommended thresholds.

Multi-platform implementation validates practical efficiency with FPGA as the primary target platform: synthesis on Xilinx Artix-7 achieves 261~Mbps throughput at 100~MHz consuming only 380 LUT resources, with the 3-regular graph topology mapping directly to LUT4 primitives and enabling fully parallel single-cycle $F_\mathrm{core}$ evaluation. ARM Cortex-M4F software execution requires 25.8~KB Flash and completes encryption in 1.66~ms, confirming feasibility across the resource-constrained IoT spectrum. Indicative ASIC synthesis on a 45~nm library yields 5.52~kGE from the FPGA-oriented RTL description, representing a conservative upper bound; dedicated ASIC restructuring through bit-serialization would reduce this figure substantially. These results establish expander-based design as a viable methodology for lightweight cryptography, offering a formally grounded design space and opening research directions in adaptive topologies, keyed graph structures, and rigorous security proofs connecting spectral properties to cryptographic resistance.
\end{abstract}

		
	\end{frontmatter}
	
	\section{Introduction}
	
	The proliferation of resource-constrained embedded systems--IoT devices, wearables, medical implants, automotive electronics--demands cryptographic primitives that deliver robust security within stringent hardware and energy budgets \cite{khan2020lightweight, lonzetta2018security, manifavas2013lightweight, yilmaz2021assure}. Traditional block ciphers like AES, while secure, impose significant overhead on devices with limited logic gates and tight power constraints \cite{dogan2014analyzing}. This has motivated two decades of lightweight cryptography research, yielding specialized designs such as PRESENT, SIMON, SPECK, GIFT, and SKINNY that optimize substitution-permutation networks (SPNs), Feistel structures, or ARX constructions through aggressive parameter tuning to minimize resource consumption \cite{bogdanov2007present, kolbl2015observations, beaulieu2015simon, cao2019related, banik2017gift, beierle2016skinny}.
	
	Despite this progress, the lightweight cipher landscape faces a critical impasse: most recent proposals represent evolutionary refinements rather than fundamentally new approaches. Incremental modifications, adjusting S-box polynomials, rearranging diffusion layers, yield diminishing returns and face increasingly skeptical scrutiny \cite{cho2010linear, biryukov2014differential}. As quantum computing advances and new cryptanalytic techniques emerge, the community requires architectures grounded in distinct mathematical principles that offer fresh sources of security and efficiency beyond optimized instances of known constructions \cite{joseph2022transitioning, dam2023survey}.
	
	This work introduces a fundamentally new design paradigm: \emph{expander-graph interaction networks}. Classical ciphers derive security from component complexity, algebraically rich S-boxes, complex modular arithmetic, or intricate round functions. Our approach derives security from \emph{structural expansion}: simple local nonlinear rules executed over a sparse, high-expansion graph. Expander graphs, combinatorial structures with strong connectivity and rapid mixing behavior, have proven powerful in theoretical computer science for derandomization, error correction, and network design, yet their cryptographic potential remains largely unexplored. By embedding lightweight Boolean functions at graph vertices, where each vertex updates based solely on its local neighborhood, we obtain diffusion that is simultaneously hardware-efficient, formally analyzable via spectral graph theory, and structurally distinct from existing cipher families.
	
	Expander-driven diffusion fundamentally differs from traditional mechanisms. SPNs require large matrix multiplications for global mixing, introducing substantial hardware cost. Feistel networks accumulate diffusion gradually through repeated complex functions. ARX designs introduce carry-chain delays and platform-dependent performance. Expander-based diffusion achieves rapid global mixing through \emph{sparse local interactions}: each bit depends on only three neighbors, yet expansion properties guarantee perturbations propagate globally in logarithmic time. Security emerges not from component strength but from composition of minimal nonlinear rules over provably well-connected topology.
	
	We instantiate this philosophy in ExpanderGraph-128 (EGC128), a 128-bit block cipher employing a balanced Feistel network over 20 rounds. Each round applies a 64-bit transformation derived from a 3-regular expander graph on 64 vertices, where each vertex executes a compact 4-input Boolean function selected to maximize nonlinearity and minimize differential uniformity.

	Beyond a single cipher design, this work establishes a new methodology:
	
	\begin{itemize}
		
	\item \textbf{Paradigm.} Expander-graph interaction networks as a fundamentally new block-cipher design approach, distinct from Goldreich-type one-way function constructions, where sparse high-expansion graphs serve as the primary diffusion mechanism and security accumulates through iterated keyed confusion--diffusion rather than single-pass hardness assumptions.
	
	\item \textbf{Theory.} Formalization connecting spectral gap, degree regularity, and long-range graph connectivity to cryptographic diffusion, showing how mixing time bounds and local differential uniformity compose with graph expansion to yield global resistance, validated through comparative MILP sensitivity analysis.
	
	\item \textbf{Instantiation.} Complete specification of EGC128 including 4-input Boolean function selection ($\mathrm{NL}=4$, degree~3), 3-regular graph construction, and balanced Feistel integration providing invertibility without $F_\mathrm{core}$ inversion.
	
	\item \textbf{Security.} MILP-proven differential bounds through 10~rounds (147.3~bits, certified optimal via SCIP; conservatively extrapolating to 413~bits for 20~rounds), formal linear trail bounds ($\geq 2^{145}$ complexity), related-key resistance with no free rounds, structural attack immunity via algebraic degree saturation and 1{,}800 affine subspace tests, and NIST SP~800-22 validation across $10^8$~bits.
	
	\item \textbf{Implementation.} FPGA synthesis on Xilinx Artix-7 as the primary target--380 LUT utilization, 261~Mbps at 100~MHz--with the 3-regular topology mapping directly to LUT4 primitives. ARM Cortex-M4F execution at 77~Kbps confirms IoT feasibility. Indicative ASIC synthesis yields 5.52~kGE in 45~nm from FPGA-oriented RTL, representing a conservative upper bound.
	
	\item \textbf{Open problems.} Formal theorems connecting spectral gap to provable security bounds, alternative expander families, keyed graph topologies, serialized ASIC-optimized variants, and side-channel countermeasures.
	
	\end{itemize}

	The rest of the paper is organized as follows. Section~2 surveys related work. Section~3 establishes theoretical foundations. Section~4 specifies EGC128. Section~5 details security analysis. Section~6 reports implementations. Section~7 discusses positioning and future directions. Section~8 concludes.
	
	\section{Related Work}
	\label{sec:related_work}
	
	Symmetric-key cryptography has evolved through distinct design paradigms grounded in different mathematical principles. We survey classical cipher approaches, examine graph-theoretic foundations enabling our methodology, and position expander-based designs within this landscape.
	
	\subsection{Classical Cipher Design Paradigms}
	
	\subsubsection{Substitution-Permutation Networks}
	
	Substitution-permutation networks (SPNs) dominate modern block cipher design, exemplified by AES and its lightweight variants. Security derives from alternating nonlinear substitution layers (S-boxes) with linear diffusion layers (permutations or matrix multiplications). PRESENT pioneered the lightweight SPN approach with a 4-bit S-box and bit permutation achieving 1,570 gate equivalents \cite{bogdanov2007present}. GIFT further optimizes this structure through refined S-box selection and permutation design, reducing hardware footprint while maintaining security margins \cite{banik2017gift}.
	
	For resource-constrained implementations, the S-box often dominates area and delay. Canright achieves compact AES S-boxes through subfield decomposition and optimized basis-change matrices \cite{Canright2005CompactSBox}, while Satoh et al.\ develop merged composite-field datapaths for minimal Rijndael implementations \cite{Satoh2001CompactAES}. Despite these optimizations, SPNs fundamentally require either large S-boxes (8 bits for strong nonlinearity) or matrix-based diffusion (MDS codes), both introducing substantial hardware cost.
	
	\subsubsection{Feistel Structures and ARX Constructions}
	
	Feistel networks achieve invertibility without inverse operations by applying nonlinear functions to half the state and XORing with the other half. Lightweight Feistel ciphers like SIMON and SPECK employ simple round functions based on bitwise rotations and AND operations \cite{beaulieu2015simon}, achieving extreme compactness at the cost of requiring many rounds for adequate diffusion. LBlock and Twine demonstrate that carefully designed Feistel structures with compact S-boxes can achieve security comparable to SPNs with competitive hardware efficiency \cite{kolbl2015observations}.
	
	ARX (Addition-Rotation-XOR) constructions leverage modular addition's nonlinearity combined with bit rotations for diffusion. SPECK exemplifies this approach, achieving minimal software implementations through operations native to general-purpose processors. However, ARX designs face platform-dependent performance: carry-chain delays impact hardware implementations, and rotation amounts must be carefully chosen to ensure adequate diffusion across rounds.
	
	\subsubsection{Limitations of Incremental Optimization}
	
	Contemporary lightweight cipher research largely focuses on evolutionary refinements: adjusting S-box algebraic properties, optimizing permutation patterns, or balancing Feistel round functions. Carlet's comprehensive treatment establishes analytical frameworks for cryptographic Boolean functions--nonlinearity, algebraic immunity, correlation immunity, differential uniformity--that guide S-box selection \cite{Carlet2007BooleanFunctions}. Nyberg's theory of differentially uniform mappings formalizes how round-key independence yields multiplicative differential probability bounds \cite{Nyberg1994DifferentiallyUniform}.
	
	While these frameworks enable systematic optimization, they reveal fundamental trade-offs: achieving high nonlinearity requires large S-boxes (increased area), strong diffusion requires dense linear layers (increased delay), and security margins require many rounds (reduced throughput). The community faces diminishing returns from parameter tuning within established paradigms, motivating exploration of architectures grounded in fundamentally different mathematical principles.
	
	\subsection{Expander Graphs: Spectral Theory and Mixing}
	
	Expander graphs, sparse regular graphs with strong connectivity properties, provide a mathematically distinct foundation for diffusion mechanisms. Unlike classical cipher components optimized for algebraic complexity, expanders derive their power from structural properties quantifiable through spectral graph theory.
	
	\subsubsection{Theoretical Foundations}
	
	Hoory, Linial, and Wigderson establish the comprehensive theoretical framework connecting edge expansion, vertex expansion, and spectral properties \cite{HooryLinialWigderson2006}. The spectral gap, the separation between the largest and second-largest eigenvalues of the adjacency matrix, determines mixing rate: larger gaps guarantee faster convergence to uniform distribution under random walks. The Expander Mixing Lemma quantifies this relationship, proving that edges distribute nearly uniformly across vertex subsets, ensuring localized perturbations propagate globally in logarithmic time.
	
	Alon establishes tight connections between eigenvalue separation and combinatorial expansion, proving that strong expansion can be verified through adjacency matrix spectrum alone \cite{Alon1986EigenvaluesExpanders}. Lubotzky, Phillips, and Sarnak identify Ramanujan graphs as optimal spectral expanders, achieving the smallest possible second eigenvalue for $(p+1)$-regular graphs \cite{Lubotzky1988Ramanujan}. These results establish theoretical limits and provide explicit construction families with provable expansion properties.
	
	Sinclair relates mixing rates to conductance and random walk convergence time through eigenvalue bounds \cite{Sinclair1992MixingFlow}, while Chapman, Linial, and Peled demonstrate constructions simultaneously guaranteeing both local neighborhood expansion and global spectral properties \cite{Chapman2019LocalGlobalExpanders}. This rich theoretical foundation provides tools for constructing graphs with quantifiable, provable mixing guarantees, properties traditionally absent from ad-hoc diffusion layer designs.
	
	\subsubsection{Cryptographic Potential}
	
	Expander graphs appear in cryptographic contexts primarily as auxiliary structures rather than core primitives. Applications include randomness extraction \cite{vadhan2012pseudorandomness}, error-correcting codes \cite{spielman1996expander}, and network constructions for secure multi-party computation. However, their potential as primary diffusion mechanisms for symmetric-key primitives remains largely unexplored.
	
	\subsection{Expander-Based Cryptographic Constructions}
	
	\subsubsection{Theoretical Foundations}
	
	Goldreich pioneers the use of expanders for constructing candidate one-way functions, demonstrating how cryptographic hardness can emerge from small local predicates combined with expander-determined neighborhoods \cite{Goldreich2000ExpanderOWF}. Critically, inversion hardness derives from expansion properties rather than predicate algebraic complexity, establishing proof-of-concept that global structural properties, not individual component sophistication, can provide security. Subsequent work explores this paradigm for pseudorandom generators \cite{applebaum2016fast}, secure computation protocols \cite{hemenway2018efficient}, and hash function constructions \cite{jao2009expander}.
	
	Nisan and Wigderson characterize $NC^{0}$ circuit limitations, proving constant-locality Boolean functions cannot create long-range dependencies or propagate correlations across large input sets \cite{NisanWigdersonNC0}. Their impossibility result validates designs combining simple Boolean rules with high-expansion topologies: diffusion provably unattainable through local operations alone becomes achievable when composed over carefully selected graph structures.
	
	Maurer and Massey introduce perfect local randomness, establishing coding-theoretic criteria for small output subsets behaving indistinguishably from uniform distributions \cite{MaurerMassey1989PLR}. Both works underscore the principle of combining simple local operations with structural mechanisms that amplify diffusion, aligned with expander-driven design philosophy.
	
	\subsubsection{Gap in Practical Block Cipher Design}
	
	Despite theoretical promise, no prior work exploits expander graphs as the primary diffusion mechanism for practical symmetric-key block ciphers. Goldreich's constructions address one-way function candidates but not block cipher requirements: invertibility, efficient key scheduling, resistance to differential and linear cryptanalysis, and practical implementation across diverse hardware platforms. Theoretical results establish feasibility of security from expansion properties but provide no concrete instantiations, security analysis methodologies, or implementation validation.
	
	The lightweight cipher community continues refining classical paradigms, SPNs, Feistel networks, ARX constructions, through incremental optimization. Graph-theoretic literature provides rich tools for constructing rapidly mixing structures with provable properties. However, these domains remain disconnected: cryptographers lack methodologies for leveraging graph expansion in cipher design, while graph theorists do not address symmetric-key security requirements.
		
	\subsubsection{Relation to Goldreich-Type Constructions}
	\label{sec:goldreich_distinction}
	
	The use of expander graphs with local Boolean functions in this work invites comparison with Goldreich's candidate one-way functions~\cite{Goldreich2000ExpanderOWF} and the pseudorandom generators of Applebaum and Raykov~\cite{applebaum2016fast}. While EGC128 draws conceptual inspiration from that paradigm, the differences in cryptographic goal, security model, and design methodology are fundamental.
	
	Goldreich constructs a candidate one-way function $G: \{0,1\}^n \to \{0,1\}^m$ by applying a fixed local predicate to expander-determined neighborhoods of an input seed in a single pass. Security rests on a computational hardness assumption for a specific (graph, predicate) pair, analyzed asymptotically ($n \to \infty$). Key material, round structure, and invertibility play no role; decryption is not a requirement.
	
	EGC128 is a keyed, invertible, iterated block cipher satisfying Shannon's confusion--diffusion paradigm across 20 rounds. Security is validated through concrete cryptanalysis, MILP-based differential and linear bounds, related-key analysis, SMT verification, and empirical testing, rather than through hardness assumptions. The construction must satisfy requirements entirely absent from Goldreich's framework: exact invertibility via the Feistel structure, a key schedule propagating 128-bit entropy across 20 independent round keys, resistance to differential and linear cryptanalysis with concrete security margins, and practical implementation on FPGA, ASIC, and microcontroller platforms.
	
	Table~\ref{tab:goldreich_vs_egc} formalises the comparison. The expander graph serves a fundamentally different role: in Goldreich's construction, the graph is the security mechanism through single-shot expansion; in EGC128, the graph provides diffusion structure within an iterated confusion--diffusion design, with security accumulating across rounds through key mixing. EGC128 is not an instantiation of Goldreich's paradigm applied to block ciphers; it is a new cipher design methodology that borrows the expander graph as a diffusion primitive and validates security through standard cryptanalytic techniques.
			
			\begin{table}[!htb]
				\centering
				\caption{Comparison of Goldreich-type constructions with EGC128. Both use expander graphs with local Boolean functions but differ fundamentally in cryptographic goal, security model, and design requirements.}
				\label{tab:goldreich_vs_egc}
				\begin{tabular}{@{}lll@{}}
					\toprule
					\textbf{Property} & \textbf{Goldreich PRG/OWF} & \textbf{EGC128 (this work)} \\
					\midrule
					Cryptographic goal  & One-way function / PRG  & Keyed block cipher \\
					Security model      & Hardness assumption      & Concrete cryptanalysis \\
					Construction type   & Single-pass              & 20-round iteration \\
					Invertibility       & Not required             & Required (Feistel) \\
					Key schedule        & None                     & 20 round keys (LFSR) \\
					Expander role       & Security mechanism       & Diffusion mechanism \\
					Parameters          & Asymptotic ($n\to\infty$)& Concrete (128-bit) \\
					Attack model        & Worst-case hardness      & Differential, linear, related-key \\
					Hardware efficiency & Not addressed            & FPGA, ASIC, MCU \\
					\bottomrule
				\end{tabular}
			\end{table}

		Table~\ref{tab:paradigm_comparison} contrasts established paradigms with our expander-based approach, highlighting the fundamental shift from component optimization to structural property exploitation.
		
		\begin{table}[!htb]
			\centering
			\caption{Comparison of block cipher design paradigms. Expander-based designs derive security from graph-theoretic structural properties rather than component algebraic complexity.}
			\label{tab:paradigm_comparison}
			\begin{tabular}{@{}llll@{}}
				\toprule
				\textbf{Paradigm} & \textbf{Diffusion} & \textbf{Nonlinearity} & \textbf{Security Basis} \\
				\midrule
				SPN & MDS matrices / bit perm. & Large S-boxes (8+ bits) & Algebraic complexity \\
				Feistel & Branch swapping & Complex F-function & Iterative composition \\
				ARX & Modular addition / rotation & Carry propagation & Integer arithmetic \\
				\textbf{Expander (this work)} & \textbf{Graph topology} & \textbf{Local rules (4-input)} & \textbf{Structural expansion} \\
				\bottomrule
			\end{tabular}
		\end{table}
		
		Rather than incremental refinement of existing paradigms, this work establishes expander-graph interaction networks as a distinct cipher design methodology, opening research directions in adaptive topologies, higher-degree local rules, keyed graph structures, and formal security proofs from spectral properties.
		
		\section{Theoretical Foundations}
		\label{sec:theoretical_foundations}
		
		We establish the mathematical principles underlying expander-based cipher design, connecting graph-theoretic expansion to cryptographic diffusion. Rather than deriving security from algebraic complexity of individual components, we exploit structural properties of sparse interaction networks.
		
		\subsection{Expander Graphs and Mixing Properties}
		
		A $d$-regular expander graph embeds every vertex in an identical local neighborhood of size $d$ while maintaining globally robust connectivity with short path lengths. The spectral gap $\lambda = 1 - |\lambda_2|$, where $\lambda_2$ denotes the second-largest eigenvalue magnitude of the normalized adjacency matrix, quantifies expansion quality. Larger gaps yield faster mixing: localized perturbations spread throughout the vertex set in $O(\log n)$ steps.
		
		The mixing time $\tau_{\text{mix}}$, bounded by
		\begin{equation}
			\tau_{\text{mix}} \leq \frac{\log n}{\lambda},
			\label{eq:mixing_time}
		\end{equation}
		measures random walk convergence to the stationary distribution. In cipher design, this translates directly to rounds required for a single-bit difference to affect the entire state with near-uniform probability. Unlike classical diffusion relying on engineered linear layers, expander-driven mixing achieves global propagation through sparse local interactions governed by spectral properties.
		
		We construct a 3-regular graph $G = (V, E)$ on $|V| = 64$ vertices for the 64-bit nonlinear layer. Each vertex $i \in \{0, 1, \ldots, 63\}$ connects to three neighbors:
		\begin{align}
			n_1(i) &= (i - 1) \bmod 64, \label{eq:neighbor1} \\
			n_2(i) &= (i + 1) \bmod 64, \label{eq:neighbor2} \\
			n_3(i) &= (i + 16) \bmod 64. \label{eq:neighbor3}
		\end{align}
		
		\begin{figure}[!htb]
			\centering
			\includegraphics[width=0.7\textwidth]{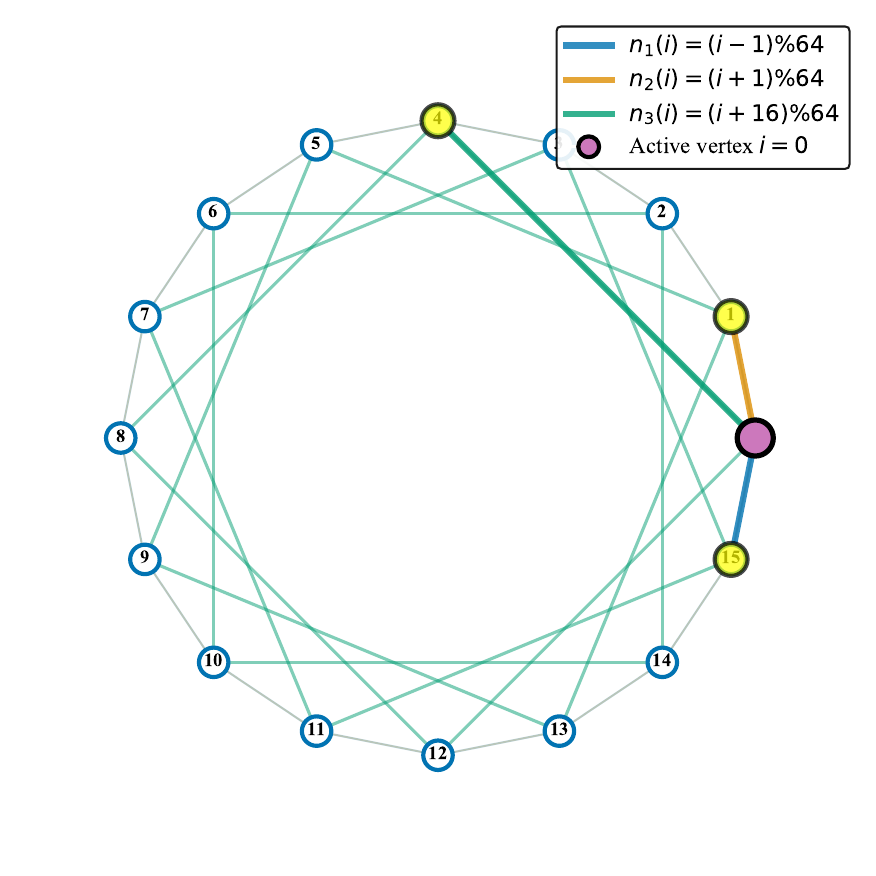}
			\caption{3-regular expander graph topology for the $F_{\text{core}}$ layer (16-vertex subset shown; full cipher uses 64 vertices). Each vertex $i$ connects to three neighbors: $n_1(i) = (i-1) \bmod 64$ (blue), $n_2(i) = (i+1) \bmod 64$ (orange), and $n_3(i) = (i+16) \bmod 64$ (green). Example: vertex 0 (magenta) receives inputs from neighbors 15, 1, and 4 (yellow). Sparse connectivity enables efficient hardware implementation while maintaining rapid mixing through expansion properties. The full cipher employs the identical topology pattern scaled to 64 vertices.}
			\label{fig:expander_topology}
		\end{figure}
		
		Figure~\ref{fig:expander_topology} illustrates the 3-regular expander graph structure governing the $F_{\text{core}}$ transformation, where each vertex maintains exactly three connections, two local neighbors at positions $(i \pm 1) \bmod 64$ and one medium-range neighbor at $(i+16) \bmod 64$, creating sparse connectivity that enables both hardware efficiency and rapid information diffusion through graph-theoretic expansion properties.
		
		This topology combines short-range connectivity ($\pm 1$ offsets) with medium-range jumps ($+16$ offset), yielding graph diameter $d_G = 8$ and average path length below 5. While not achieving Ramanujan optimality, this graph represents a deliberate trade-off favoring implementation simplicity over theoretical optimality. Neighbor computations reduce to bit rotations and XOR operations, eliminating complex wiring or lookup tables. The spectral gap remains sufficient to guarantee logarithmic mixing time while maintaining predictable routing costs across platforms.
		
		Degree $d = 3$ represents the minimum enabling sublinear mixing while preserving sparsity: each vertex update requires only four input bits (self plus three neighbors), mapping directly to FPGA LUT4 primitives. Degree-2 graphs form simple cycles with linear mixing time $\tau_{\text{mix}} = \Theta(n^2)$ unsuitable for cryptography. Higher degrees would accelerate mixing but increase hardware cost proportionally.
		
		\subsection{Local Boolean Functions for Cryptographic Strength}
		
		We embed a carefully selected 4-input Boolean function at each vertex, executing identical update rules across the entire state. Our function $f: \{0,1\}^4 \rightarrow \{0,1\}$, designated Rule-A with truth table \texttt{0x036F}, achieves optimal cryptographic properties within the 4-variable constraint space.
		
		We identified Rule-A through exhaustive enumeration of all $2^{16} = 65{,}536$ possible 4-input Boolean functions, applying the following criteria:
		
		\begin{enumerate}
			\item \textbf{Balance}: Exactly 8 zeros and 8 ones, ensuring unbiased output distribution.
			\item \textbf{Maximum nonlinearity}: $\text{NL}(f) = 4$, the theoretical maximum for balanced 4-variable functions.
			\item \textbf{Minimum differential uniformity}: $\text{DU}(f) = 12$, yielding worst-case single-node probability $p_{\max} = 3/4$.
			\item \textbf{Algebraic degree}: $\deg(f) = 3$ provides resistance to algebraic attacks with efficient gate realization.
		\end{enumerate}
		
		Exactly 4,158 functions satisfied balance, maximum nonlinearity, and degree-3 constraints. Among these, Rule-A minimizes differential uniformity with algebraic normal form:
		\begin{equation}
			f(x_0, x_1, x_2, x_3) = 1 \oplus x_2 \oplus x_0 x_2 \oplus x_1 x_2 \oplus x_1 x_3 \oplus x_0 x_2 x_3.
			\label{eq:rule_a_anf}
		\end{equation}
		
		This ANF contains six monomials with extensive subexpression sharing, enabling hardware synthesis to extract common logic and minimize gate count. On FPGAs, the function maps to a single LUT4 primitive; in ASICs, synthesis achieves fewer than 5 equivalent gates per vertex.
		
		The local differential probability $p_{\max} = 3/4$ combines with graph expansion to yield strong global resistance. A single input difference propagates to $\Theta(\log n)$ vertices within constant rounds, and the Feistel structure ensures differences traverse the graph layer twice per full round. MILP analysis (Section~5) confirms super-linear growth in active nodes, validating that local differential uniformity composes favorably with expansion-driven activation.
		
		\subsection{Connecting Expansion to Security}
		
		We formalize how graph-theoretic expansion translates to cryptographic security through three observations linking local properties to global resistance.
		
		\begin{observation}[Activation Growth]
			\label{obs:activation}
			Let $A_r \subseteq V$ denote active vertices in round $r$ under truncated differential model. For any initial difference with $|A_0| \geq 1$, the expander property guarantees
			\begin{equation}
				|A_{r+1}| \geq \min\bigl(|V|, \, c \cdot |A_r|\bigr)
			\end{equation}
			for expansion constant $c > 1$ determined by spectral gap and neighbor structure.
		\end{observation}
		
		\begin{observation}[Differential Weight Accumulation]
			\label{obs:weight}
			Each active vertex contributes differential weight $w_{\text{node}} = -\log_2(3/4) \approx 0.415$ bits. Combining Observation~\ref{obs:activation} with weight accumulation yields minimum differential weight:
			\begin{equation}
				W(r) \geq w_{\text{node}} \cdot \sum_{i=0}^{r-1} |A_i|.
				\label{eq:weight_bound}
			\end{equation}
		\end{observation}
		
		\begin{observation}[Empirical Validation]
			\label{obs:milp}
			MILP-based truncated differential analysis confirms minimum active node counts: $\{4, 13, 29, 53\}$ for rounds $\{1, 2, 3, 4\}$ respectively. Growth rates $\{3.25, 2.23, 1.83\}$ demonstrate diminishing expansion as saturation approaches, consistent with bounded vertex set $|V| = 64$.
		\end{observation}
		
		These observations establish that security emerges from structural composition: simple local rules with moderate individual resistance combine with provably expanding topology to yield global transformations resisting differential cryptanalysis. The Feistel structure amplifies this by forcing every difference to traverse the graph-based F-function, creating cascading activation patterns.
		
		\subsection{Design Rationale and Parameter Selection}
		
		\paragraph{Twenty-Round Feistel Structure} Equation~\eqref{eq:weight_bound} and Observation~\ref{obs:milp} suggest 9--11 rounds suffice for 64--80 bits of differential weight. We adopt 20 rounds providing 2× security margin, accommodating cryptanalytic advances and ensuring long-term robustness.
		
		\paragraph{Graph Topology Selection} The 3-regular $\pm 1, +16$ topology balances expansion quality against implementation cost. Ramanujan graphs achieve optimal spectral gaps but require irregular neighbor patterns or lookup-based routing. Our regular construction enables neighbor computation through modular arithmetic (bit rotations), yielding predictable routing. The $+16$ offset ensures local neighborhoods span full state width within 4 graph applications, accelerating global mixing.
		
		\paragraph{Local Function Complexity} Four-input Boolean functions represent minimum complexity enabling meaningful cryptographic properties (maximum nonlinearity 4, nontrivial differential uniformity) while mapping to standard hardware primitives. Three-input functions achieve only $\text{NL} = 2$ and lack design space; five-input functions would require LUT5 or multi-level logic, increasing area and delay.
		
		\paragraph{Key Schedule Design} The LFSR-based key schedule employs primitive polynomial $x^{64} + x^4 + x^3 + x + 1$, guaranteeing period $2^{64} - 1$ and ensuring full key entropy propagates across rounds. Round constants derived from $\pi$ digits eliminate residual symmetries. This lightweight approach maintains consistency with deriving security from structural properties rather than component complexity.
		
		\subsection{Design Space and Generalizability}
		
		Our approach instantiates one point in a rich design space amenable to systematic exploration. Alternative expander families (Ramanujan graphs, Cayley graphs, random $d$-regular graphs) offer different expansion-versus-structure trade-offs potentially reducing round counts or improving resistance bounds. Higher-degree graphs ($d = 4, 5$) accelerate mixing at increased hardware cost, while larger state sizes ($n = 128, 256$ vertices) extend block length preserving logarithmic mixing time. Adaptive or keyed topologies where graph structure depends on key material introduce security through topology variation, and local function diversity across vertices could disrupt structured attacks while maintaining regularity. The design's foundation in graph-theoretic rather than number-theoretic hardness positions it favorably for post-quantum security. These directions establish expander-based design as a \emph{methodology} for constructing lightweight primitives grounded in combinatorial mathematics, providing tools for systematic design space exploration of future expander-based constructions.
		
		\section{ExpanderGraph-128 Specification}
		\label{sec:eg_specification}
		
		We present the complete algorithmic specification of EGC128, demonstrating how the theoretical framework from Section~3 translates to a concrete cipher design. The specification emphasizes the direct mapping between graph-theoretic constructs (vertex updates, neighbor connectivity) and cryptographic operations (bit transformations, state mixing).
		
		\subsection{Block Cipher Structure}
		
		EGC128 operates on 128-bit blocks using 128-bit keys through a 20-round balanced Feistel network. We parse the 128-bit plaintext $P$ and key $K$ as follows:
		\begin{align}
			P &= L_0 \| R_0 \quad \text{where } L_0, R_0 \in \{0,1\}^{64}, \label{eq:plaintext_parse} \\
			K &= K_{\text{high}} \| K_{\text{low}} \quad \text{where } K_{\text{high}}, K_{\text{low}} \in \{0,1\}^{64}. \label{eq:key_parse}
		\end{align}
		
		Each Feistel round $r \in \{0, 1, \ldots, 19\}$ applies the expander-graph transformation $F_{\text{core}}$ to the right branch, combines the result with round key $RK_r$, and swaps branches:
		\begin{align}
			L_{r+1} &= R_r, \label{eq:feistel_left} \\
			R_{r+1} &= L_r \oplus F_{\text{core}}(R_r) \oplus RK_r. \label{eq:feistel_right}
		\end{align}
		
		After 20 rounds, the ciphertext emerges as $C = L_{20} \| R_{20}$. Decryption applies identical operations with round keys in reverse order $\{RK_{19}, RK_{18}, \ldots, RK_0\}$, exploiting the Feistel structure's inherent invertibility without requiring $F_{\text{core}}$ inversion.

		\begin{figure}[!htb]
			\centering
			\includegraphics[width=0.85\textwidth]{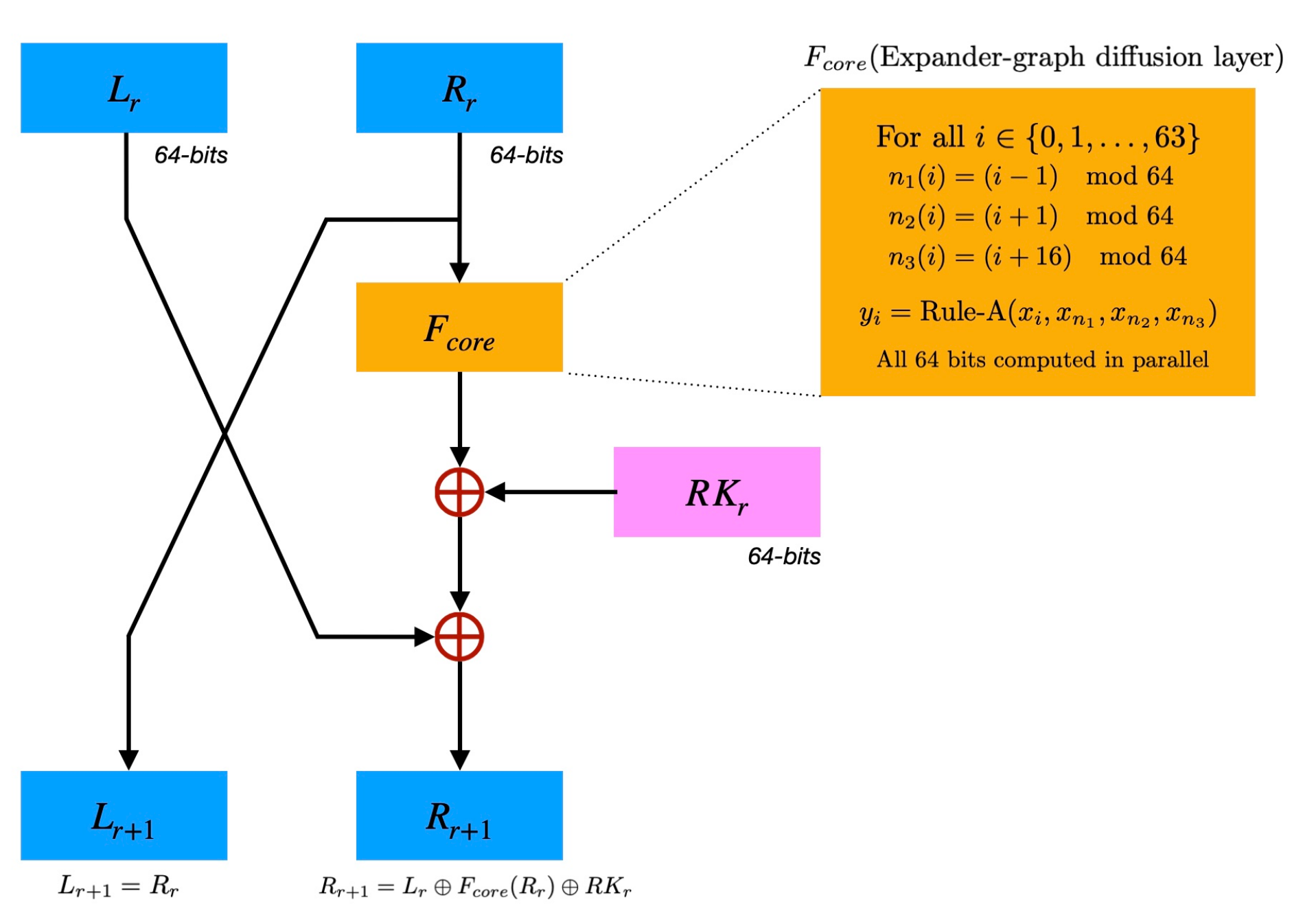}
			\caption{Single Feistel round architecture. The 128-bit state $(L_r, R_r)$ transforms to $(L_{r+1}, R_{r+1})$ through branch swap and expander-graph-based nonlinear function $F_{\text{core}}$. The right branch $R_r$ undergoes graph-theoretic diffusion where each of 64 bits applies Rule-A using inputs from three neighbors determined by the 3-regular topology. Round key $RK_r$ combines with the transformed branch via XOR operations to produce $R_{r+1}$, while $L_{r+1}$ receives the unmodified right branch, ensuring invertibility.}
			\label{fig:round_function}
		\end{figure}
		
		The complete dataflow for a single Feistel round is depicted in Figure~\ref{fig:round_function}, showing how the right branch $R_r$ undergoes expander-graph transformation via $F_{\text{core}}$, combines with round key $RK_r$ through XOR operations, and merges with the left branch $L_r$ to produce the next state $(L_{r+1}, R_{r+1})$, while the branch swap ensures invertibility essential for decryption.
		
		\subsection{Expander-Graph Nonlinear Transformation}
		
		The function $F_{\text{core}}: \{0,1\}^{64} \rightarrow \{0,1\}^{64}$ implements the 3-regular expander graph interaction layer, embodying the core design principle that security emerges from sparse structural connectivity rather than component complexity. For input $X = (x_0, x_1, \ldots, x_{63})$, each output bit $y_i$ results from applying Rule-A to vertex $i$ and its three neighbors:
		\begin{equation}
			y_i = f(x_i, x_{n_1(i)}, x_{n_2(i)}, x_{n_3(i)}), \quad i \in \{0, 1, \ldots, 63\},
			\label{eq:fcore_bit}
		\end{equation}
		where neighbor indices follow the topology established in Section~3.1:
		\begin{align}
			n_1(i) &= (i - 1) \bmod 64, \label{eq:spec_neighbor1} \\
			n_2(i) &= (i + 1) \bmod 64, \label{eq:spec_neighbor2} \\
			n_3(i) &= (i + 16) \bmod 64. \label{eq:spec_neighbor3}
		\end{align}
		
		Rule-A, specified by truth table \texttt{0x036F} and ANF given in Equation~\eqref{eq:rule_a_anf}, computes each bit transformation. The 64 bit updates execute independently in parallel, each depends only on four input bits, enabling full hardware parallelization without data hazards or ordering constraints.
		
		\subsection{Key Schedule}
		
		Round keys derive from a lightweight 64-bit LFSR initialized with the upper key half, combined with the lower key half and fixed round constants. We initialize LFSR state $S_0$ as:
		\begin{equation}
			S_0 = \begin{cases}
				K_{\text{high}} & \text{if } K_{\text{high}} \neq 0, \\
				1 & \text{if } K_{\text{high}} = 0,
			\end{cases}
			\label{eq:lfsr_init}
		\end{equation}
		where the all-zero state receives special handling to ensure maximal LFSR period $2^{64} - 1$.
		
		\begin{figure}[!htb]
			\centering
			\includegraphics[width=0.9\textwidth]{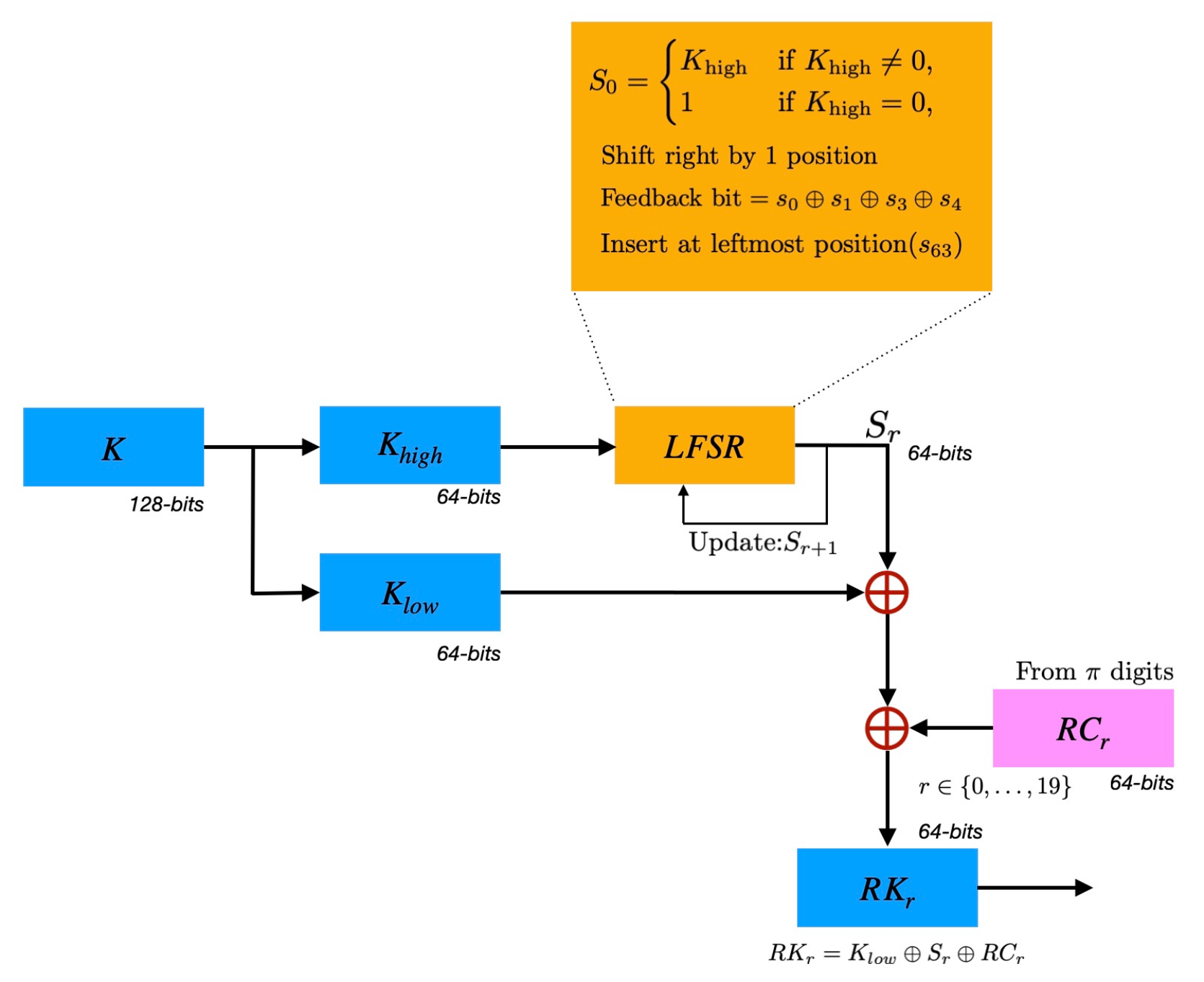}
			\caption{Key schedule architecture for 20-round ExpanderGraph-128. The 128-bit master key $K$ splits into $K_{\text{high}}$ and $K_{\text{low}}$. The upper 64 bits initialize a maximal-period LFSR with primitive polynomial $x^{64} + x^4 + x^3 + x + 1$, generating evolving state $S_r$ through shift-and-XOR operations with feedback taps at positions $\{0, 1, 3, 4\}$. Round key $RK_r$ derives from three-way XOR: the constant lower key half $K_{\text{low}}$, current LFSR state $S_r$, and round-specific constant $RC_r$ extracted from $\pi$ digits. This lightweight design requires no S-box lookups while ensuring unique keys across all rounds.}
			\label{fig:key_schedule}
		\end{figure}
		
		For each round $r$, we compute:
		\begin{equation}
			RK_r = K_{\text{low}} \oplus S_r \oplus RC_r,
			\label{eq:round_key}
		\end{equation}
		where $RC_r$ denotes the fixed 64-bit round constant for round $r$, and $S_{r+1}$ follows from one LFSR update:
		\begin{equation}
			S_{r+1} = (S_r \gg 1) \, | \, \bigl((s_0 \oplus s_1 \oplus s_3 \oplus s_4) \ll 63\bigr).
			\label{eq:lfsr_step}
		\end{equation}
		
		The LFSR employs primitive polynomial $x^{64} + x^4 + x^3 + x + 1$ with taps at bit positions $\{0, 1, 3, 4\}$. Round constants $\{RC_0, RC_1, \ldots, RC_{19}\}$ derive from hexadecimal digits of $\pi$, eliminating structural symmetries:
		\begin{align*}
			RC_0 &= \texttt{0x243f6a8885a308d3}, \quad RC_1 = \texttt{0x13198a2e03707344}, \\
			RC_2 &= \texttt{0xa4093822299f31d0}, \quad \ldots \quad RC_{19} = \texttt{0x3707344a40938220}.
		\end{align*}
		
		Figure~\ref{fig:key_schedule} presents the key schedule architecture, where the 128-bit master key $K$ splits into upper and lower halves that respectively initialize a 64-bit LFSR with primitive polynomial $x^{64} + x^4 + x^3 + x + 1$ and provide a constant key component, generating unique round keys $RK_r = K_{\text{low}} \oplus S_r \oplus RC_r$ for each of the 20 rounds through lightweight shift-and-XOR operations without requiring S-box lookups or complex key-dependent permutations.
		
		This key schedule avoids lookup-table expansion (as in AES) or complex ARX mixing, maintaining consistency with the design philosophy: derive security from graph-based structural diffusion rather than key-schedule complexity.
		Algorithm~\ref{alg:encryption} specifies the complete encryption procedure, and Algorithm~\ref{alg:decryption} provides decryption.
		
		\begin{algorithm}[t]
			\caption{EGC128 Encryption}
			\label{alg:encryption}
			\begin{algorithmic}[1]
				\Require 128-bit key $K = K_{\text{high}} \| K_{\text{low}}$, plaintext $P = L_0 \| R_0$
				\Ensure 128-bit ciphertext $C$
				\State $S_0 \gets K_{\text{high}}$; \textbf{if} $S_0 = 0$ \textbf{then} $S_0 \gets 1$ \Comment{Initialize LFSR}
				\For{$r = 0, 1, \ldots, 19$}
				\State $RK_r \gets K_{\text{low}} \oplus S_r \oplus RC_r$ \Comment{Generate round key}
				\State $S_{r+1} \gets \text{LFSR update from } S_r$ via Eq.~\eqref{eq:lfsr_step}
				\State $F_{\text{out}} \gets F_{\text{core}}(R_r)$ via Eq.~\eqref{eq:fcore_bit} \Comment{Apply expander graph}
				\State $L_{r+1} \gets R_r$; \quad $R_{r+1} \gets L_r \oplus F_{\text{out}} \oplus RK_r$ \Comment{Feistel update}
				\EndFor
				\State \Return $C \gets L_{20} \| R_{20}$
			\end{algorithmic}
		\end{algorithm}
		
		\begin{algorithm}[t]
			\caption{EGC128 Decryption}
			\label{alg:decryption}
			\begin{algorithmic}[1]
				\Require 128-bit key $K = K_{\text{high}} \| K_{\text{low}}$, ciphertext $C = L_{20} \| R_{20}$
				\Ensure 128-bit plaintext $P$
				\State Generate all round keys $\{RK_0, RK_1, \ldots, RK_{19}\}$ identically to encryption
				\For{$r = 19, 18, \ldots, 0$} \Comment{Reverse round order}
				\State $R_r \gets L_{r+1}$ \Comment{Invert branch swap}
				\State $F_{\text{out}} \gets F_{\text{core}}(R_r)$ \Comment{Same $F_{\text{core}}$ as encryption}
				\State $L_r \gets R_{r+1} \oplus F_{\text{out}} \oplus RK_r$ \Comment{Invert XOR operation}
				\EndFor
				\State \Return $P \gets L_0 \| R_0$
			\end{algorithmic}
		\end{algorithm}
		
		\subsection{Implementation Considerations}
		
		We highlight key properties relevant to efficient and secure implementation. All 64 output bits of $F_{\text{core}}$ compute independently from disjoint 4-bit input subsets determined by fixed graph topology, enabling full hardware parallelization where 64 instances of Rule-A execute simultaneously without data dependencies for single-cycle evaluation in pipelined architectures. The algorithms contain no conditional branches dependent on secret data, all operations (XOR, bit selection, neighbor index computation, Boolean function evaluation) execute in constant time regardless of input values, providing algorithmic-level resistance to timing side-channel attacks. Both encryption and decryption permit precomputing all 20 round keys before block processing, allowing implementations to store $\{RK_0, \ldots, RK_{19}\}$ in registers (hardware) or fast memory (software), amortizing key-schedule cost across multiple blocks in parallelizable cipher modes. We adopt big-endian bit ordering where bit 0 denotes the least significant bit, and implementations must maintain consistent ordering across all operations to ensure cross-platform interoperability and test vector validation.
		
		\subsection{Test Vectors}
		
		Table~\ref{tab:test_vectors} provides three representative test vectors spanning boundary conditions (all-zeros, all-ones) and pseudorandom cases. The complete set of ten official vectors appears in Appendix~A.
		
		\begin{table}[h]
			\centering
			\caption{Representative EGC128 test vectors (128-bit key and block, 20 rounds). Complete vector set in Appendix~A.}
			\label{tab:test_vectors}
			\begin{tabular}{@{}lll@{}}
				\toprule
				\textbf{Vector} & \textbf{Type} & \textbf{Hexadecimal Value} \\
				\midrule
				\multirow{3}{*}{TV1} & Key & \texttt{0x00000000000000000000000000000000} \\
				& Plaintext & \texttt{0x00000000000000000000000000000000} \\
				& Ciphertext & \texttt{0x054e2db44cd3907d7c814c56070da703} \\
				\midrule
				\multirow{3}{*}{TV4} & Key & \texttt{0xffffffffffffffffffffffffffffffff} \\
				& Plaintext & \texttt{0xffffffffffffffffffffffffffffffff} \\
				& Ciphertext & \texttt{0x797644AEE6B69C4C28AC59BDCCE7FF19} \\
				\midrule
				\multirow{3}{*}{TV10} & Key & \texttt{0x3C4F1A279BD80256E1F0C3A5D4976B8E} \\
				& Plaintext & \texttt{0x9A7C3E2B10F4D8C6B5E1A2938476D0F1} \\
				& Ciphertext & \texttt{0x0C578E13690158046726B86187D850DA} \\
				\bottomrule
			\end{tabular}
		\end{table}
		
		These vectors enable straightforward implementation verification: deterministic key-plaintext pairs produce unique ciphertexts, confirming correct algorithm execution across diverse platforms (FPGA, ASIC, microcontroller) and programming environments (HDL, C, Python).
		
		\section{Security Analysis}
		\label{sec:security_analysis}
		
		We evaluate EGC128 through multiple complementary methodologies that establish resistance to classical cryptanalytic attacks. Our primary contribution is rigorous MILP-based truncated differential analysis via SCIP solver, establishing proven optimal bounds for 10 rounds and demonstrating characteristic expansion-to-saturation dynamics. This analysis reveals minimum active Rule-A counts of \{4, 13, 29, 53, 85, 125, 173, 229, 291, 355\} across rounds 1--10, with conservative extrapolation yielding differential security exceeding 400 bits for the full 20-round cipher. 
		
		Complementary analyses include SMT verification formally eliminating zero-output differentials through four rounds, empirical linear cryptanalysis via Walsh spectrum analysis providing preliminary correlation decay measurements, avalanche testing demonstrating 49\% state diffusion by round 20, and comprehensive statistical randomness evaluation via NIST SP~800-22 confirming pseudorandom output quality across $10^8$ bits. This multi-layered approach validates that security emerges from the synergy between local nonlinearity (simple 4-input Boolean functions) and global expansion properties (3-regular graph topology with strong spectral gap) established in Section~3.
		
		The differential analysis represents the strongest security foundation, as differential cryptanalysis historically provides tighter bounds than linear cryptanalysis for block ciphers. The rigorous MILP methodology, combined with supporting evidence from SMT verification and empirical testing, establishes EGC128's resistance to known classical attacks while demonstrating that graph-theoretic expansion properties translate directly into measurable cryptographic security.
		
		%% From here 
		\subsection{Primary Security Analysis: Differential Cryptanalysis}
		
		\subsubsection{MILP-Based Truncated Differential Bounds}
		
		We employ mixed-integer linear programming (MILP) to derive provable lower bounds on active Rule-A applications under truncated differential propagation. The model encodes the Feistel structure, expander-graph topology, and differential propagation rules, solving via SCIP 9.2.4 with aggressive presolve and symmetry detection.
		
		For the 64-bit $F_{\text{core}}$ function based on the 3-regular graph, we introduce binary activity variables $s_F(r,i) \in \{0,1\}$ indicating whether Rule-A at position $i$ is active (has nonzero input difference) in round $r$. The graph topology enforces:
		\begin{align}
			s_F(r,i) &\geq R_r[i], \label{eq:milp_constraint1} \\
			s_F(r,i) &\geq R_r[n_1(i)], \quad s_F(r,i) \geq R_r[n_2(i)], \quad s_F(r,i) \geq R_r[n_3(i)], \label{eq:milp_constraint2} \\
			s_F(r,i) &\leq R_r[i] + R_r[n_1(i)] + R_r[n_2(i)] + R_r[n_3(i)], \label{eq:milp_constraint3}
		\end{align}
		where neighbor indices $n_1(i) = (i-1) \bmod 64$, $n_2(i) = (i+1) \bmod 64$, $n_3(i) = (i+16) \bmod 64$ encode the sparse connectivity pattern. These constraints capture expansion-driven activation: if any input to Rule-A at position $i$ has nonzero difference, the Rule-A instance activates.
		
		Additionally, we enforce that active Rule-A instances must produce active output differences, following standard truncated differential methodology:
		\begin{equation}
			F_{\text{out}}(r,i) = s_F(r,i), \quad \forall r,i.
			\label{eq:fout_equals_sf}
		\end{equation}
		This prevents differential cancellation and ensures security bounds reflect worst-case propagation where all activated nodes contribute to output differences.
		
		The Feistel structure couples left and right branches via:
		\begin{align}
			L_{r+1}[i] &= R_r[i], \label{eq:feistel_milp1} \\
			R_{r+1}[i] &\geq L_r[i], \quad R_{r+1}[i] \geq F_{\text{out}}(r,i), \quad R_{r+1}[i] \leq L_r[i] + F_{\text{out}}(r,i), \label{eq:feistel_milp2}
		\end{align}
		with boundary conditions $\sum_{i=0}^{63} L_0[i] \geq 1$, $\sum_{i=0}^{63} R_0[i] \geq 1$ (nontrivial input in both branches), and $\sum_{i=0}^{63} (L_{N_r}[i] + R_{N_r}[i]) \geq 1$ (nontrivial output) preventing trivial all-zero characteristics.
		
		We minimize total active Rule-A count across all rounds:
		\begin{equation}
			\min \sum_{r=0}^{N_r-1} \sum_{i=0}^{63} s_F(r,i),
			\label{eq:milp_objective}
		\end{equation}
		solving for $R \in \{1,2,\ldots,10\}$ Feistel rounds. Table~\ref{tab:milp_feistel_bounds} presents proven optimal minimum active Rule-A counts.
		
		\begin{table}[!htb]
			\centering
			\caption{MILP-derived minimum active Rule-A counts for ExpanderGraph-128. All results proven optimal via SCIP 9.2.4 solver. Growth rates demonstrate initial super-linear expansion (rounds 1--4) transitioning to saturation behavior (rounds 5--10) as active nodes approach the 64-vertex graph capacity.}
			\label{tab:milp_feistel_bounds}
			\begin{tabular}{@{}cccc@{}}
				\toprule
				\textbf{Rounds $R$} & \textbf{Min Active Rule-A} & \textbf{Growth Rate} & \textbf{Differential Weight (bits)} \\
				\midrule
				1 & 4 & --- & 1.7 \\
				2 & 13 & 3.25$\times$ & 5.4 \\
				3 & 29 & 2.23$\times$ & 12.0 \\
				4 & 53 & 1.83$\times$ & 22.0 \\
				5 & 85 & 1.60$\times$ & 35.3 \\
				6 & 125 & 1.47$\times$ & 51.9 \\
				7 & 173 & 1.38$\times$ & 71.8 \\
				8 & 229 & 1.32$\times$ & 95.0 \\
				9 & 291 & 1.27$\times$ & 120.8 \\
				10 & 355 & 1.22$\times$ & 147.3 \\
				\bottomrule
			\end{tabular}
		\end{table}
		
		The results demonstrate characteristic expansion-to-saturation dynamics. Rounds 1--4 exhibit super-linear growth with rates $\{3.25, 2.23, 1.83, 1.60\}$, reflecting rapid activation cascade as differences propagate through the expander graph. Rounds 5--10 show diminishing growth rates $\{1.47, 1.38, 1.32, 1.27, 1.22\}$ as the active set approaches full graph coverage, by round 9, 62 of 64 vertices activate per round, and round 10 achieves complete saturation with all 64 vertices active.
		
		Each active Rule-A contributes differential weight $w_{\text{node}} = -\log_2(3/4) \approx 0.415$ bits, corresponding to worst-case differential probability $3/4$ per node. Combining MILP bounds with local probabilities yields minimum differential weights shown in Table~\ref{tab:milp_feistel_bounds}. The proven minimum for 10 rounds establishes $W(10) \geq 147.3$ bits, substantially exceeding the 128-bit security threshold.
		
		%\paragraph{Security Margin for 20 Rounds.}
		For the full 20-round EGC128 cipher, we observe that saturation occurs by round 10, where all 64 graph vertices activate. In saturated rounds, every Rule-A instance processes active differences, contributing the full 64 active nodes per round. Conservative extrapolation yields:
		\begin{equation}
			W(20) \geq W(10) + 10 \times 64 \times w_{\text{node}} = 147.3 + 10 \times 64 \times 0.415 \approx 413 \text{ bits}.
			\label{eq:weight_20rounds}
		\end{equation}
		
		This estimate, grounded in 10 rounds of proven optimal MILP bounds followed by saturation-based extrapolation, establishes that EGC128 provides differential security margins far exceeding practical attack feasibility. The 20-round design offers substantial margin against differential cryptanalysis, with proven bounds through half the cipher demonstrating the effectiveness of expander-driven diffusion.
		
		\subsubsection{Activation Growth Pattern Analysis}
		
		\begin{figure}[!htb]
			\centering
			\begin{subfigure}{0.48\textwidth}
				\centering
				\includegraphics[width=\linewidth]{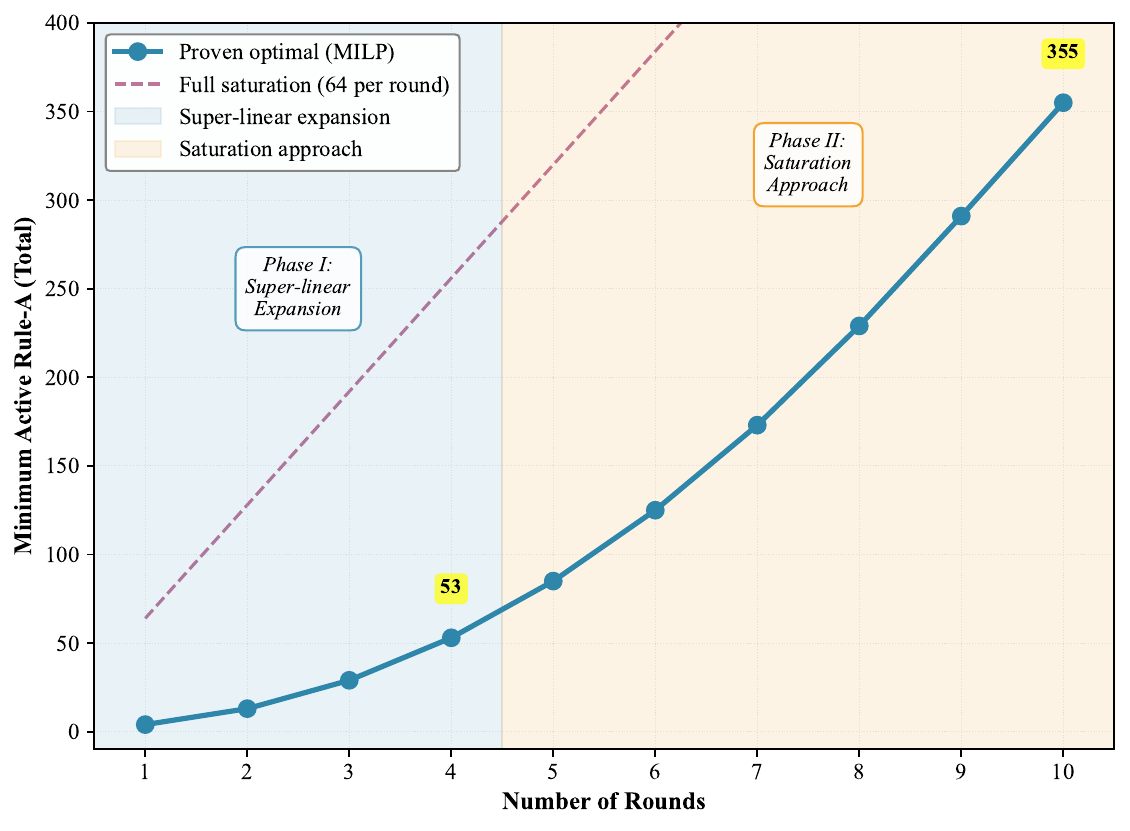}
				\caption{Cumulative Active Rule-A Growth}
				\label{fig:cumulative_growth}
			\end{subfigure}
			\hfill
			\begin{subfigure}{0.48\textwidth}
				\centering
				\includegraphics[width=\linewidth]{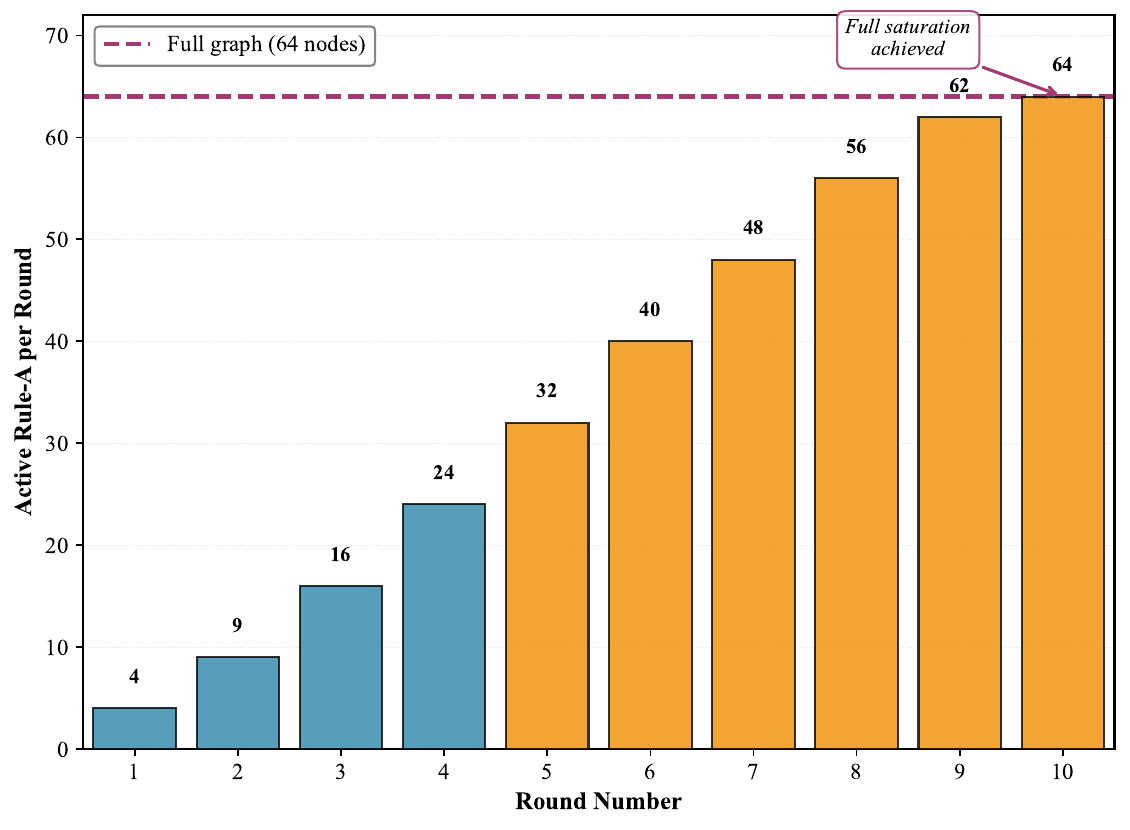}
				\caption{Per-Round Activation Increments}
				\label{fig:per_round_increments}
			\end{subfigure}
			\caption{MILP differential analysis results showing: (a) the cumulative growth of active Rule-A nodes across rounds, demonstrating two-phase behavior with super-linear expansion followed by saturation approach; and (b) the per-round activation increments, highlighting the progression toward full saturation at 64 nodes per round.}
			\label{fig:milp_growth}
		\end{figure}
		
		Figure~\ref{fig:milp_growth} illustrates the complete activation growth trajectory. The transition from super-linear expansion (rounds 1--4) to saturation approach (rounds 5--10) precisely matches theoretical predictions from spectral graph theory: the spectral gap guarantees logarithmic mixing time, producing rapid initial spread, while finite state size imposes an ultimate bound on active node count.

		The per-round activation increments reveal the saturation dynamics explicitly:
		\begin{equation}
			\Delta A(r) = A(r) - A(r-1): \quad \{4, 9, 16, 24, 32, 40, 48, 56, 62, 64\}.
			\label{eq:per_round_activation}
		\end{equation}
		Round-to-round increments grow from 4 (round 1) to 64 (round 10), with rounds 8--10 approaching and achieving full graph coverage. This progression, from localized activation to complete diffusion, validates the core design principle that sparse graph structure with simple local rules yields strong global mixing.
		
		\subsubsection{Exact Single-Layer Differential Analysis}
		
		To validate local differential behavior independent of Feistel structure, we constructed exact MILP models for the $F_{\text{core}}$ interaction layer on reduced 16-bit and 32-bit state sizes (preserving 3-regular topology with scaled neighbor offsets). Binary selector variables $\lambda_{j,(a,b)}$ choose valid differential transitions from Rule-A's difference distribution table (DDT) at each vertex $j$:
		\begin{equation}
			W = \sum_{j=0}^{n-1} \sum_{(a,b) \in \text{DDT}} \lambda_{j,(a,b)} \cdot \bigl(-\log_2(\text{DDT}_f(a,b)/16)\bigr).
			\label{eq:exact_layer_weight}
		\end{equation}
		
		Solving for both 16-bit and 32-bit instances yielded identical minimum differential weights:
		\begin{equation}
			W_{F,16}^{\min} = W_{F,32}^{\min} \approx 3.415 \text{ bits},
		\end{equation}
		confirming that optimal single-layer differentials depend only on local neighborhood structure, not global state dimensions. This dimension-independence validates the local properties of the expander-graph design and demonstrates that expansion behavior is intrinsic to the 3-regular topology.
		
		\subsubsection{Empirical Differential Testing on Full Cipher}
		
		We measured actual differential probabilities on reduced-round instances of the complete 128-bit cipher. For each input difference $\Delta P$ and round count $r \in \{1,\ldots,6\}$, we sampled 8000 random plaintext pairs $(P, P \oplus \Delta P)$ and recorded maximum empirical differential probability across all observed output differences $\Delta C = E_K^{(r)}(P) \oplus E_K^{(r)}(P \oplus \Delta P)$.
		
		Table~\ref{tab:empirical_differential} shows rapid exponential decay in maximum observed probabilities. Single-bit differences yield $\text{DP} < 2^{-8}$ by round 6, while two-bit differences achieve $\text{DP} < 2^{-11.97}$ over the same span. No anomalous high-probability differentials emerged across 56,000 total samples (8000 samples $\times$ 7 difference patterns), confirming smooth diffusion without structural weaknesses.
		
		\begin{table}[!htb]
			\centering
			\caption{Empirical maximum differential probabilities from 8000 random samples per (difference, round) pair. Input differences span single-bit and adjacent-bit-pair patterns across state boundaries.}
			\label{tab:empirical_differential}
			\begin{tabular}{@{}lccc@{}}
				\toprule
				\textbf{Input Difference $\Delta P$} & \textbf{Rounds $r$} & \textbf{Max Empirical DP} & \textbf{Differential Weight (bits)} \\
				\midrule
				Single bit (position 0) & 3 & $1.69 \times 10^{-2}$ & 5.89 \\
				Single bit (position 0) & 6 & $3.63 \times 10^{-3}$ & 8.11 \\
				Single bit (position 63) & 6 & $1.50 \times 10^{-3}$ & 9.38 \\
				Single bit (position 64) & 6 & $2.88 \times 10^{-3}$ & 8.44 \\
				Single bit (position 127) & 6 & $2.75 \times 10^{-3}$ & 8.51 \\
				Bit pair (positions 0,1) & 6 & $2.50 \times 10^{-4}$ & 11.97 \\
				Bit pair (positions 63,64) & 6 & $1.38 \times 10^{-3}$ & 9.51 \\
				\bottomrule
			\end{tabular}
		\end{table}
		
		\subsubsection{SMT-Based Verification and Impossible Differentials}
		
		To formally verify absence of unexpected low-weight differentials in early rounds, we constructed bit-precise Z3 Satisfiability Modulo Theories (SMT) models for a reduced 32-bit Feistel cipher (16-bit branches, 3-regular graph with scaled offsets $\{-1,+1,+4\}$). The reduced model preserves local structure while enabling exact satisfiability queries infeasible for 128-bit instances.
		
		For input differences $\Delta P \in \{\text{single-bit variants}, \text{two-bit adjacent pairs}\}$ and rounds $r \in \{2,3,4\}$, we queried satisfiability of zero-output constraint $\Delta C = 0$. Z3 returned \texttt{UNSAT} for all 36 tested combinations, formally proving no plaintext pair produces zero output difference for any low-weight input over 2--4 rounds. This eliminates zero-differential distinguishers in the early rounds.
		
		Extending to impossible differential search, we enumerated all single-bit output differences $\Delta C = \text{bit}_j$ for $j \in \{0,\ldots,31\}$ under the same input set and rounds $r \in \{2,3\}$. Z3 confirmed \texttt{UNSAT} for all configurations: no low-weight input ($\text{HW}(\Delta P) \leq 2$) can produce single-bit output ($\text{HW}(\Delta C) = 1$) within 2--3 rounds. This establishes rapid expansion: low-weight differences must activate $\geq 2$ output bits by round 2, providing additional evidence for expansion-driven diffusion quality.

		\subsection{Complementary Analyses}

			\subsubsection{Linear Cryptanalysis}\label{sec:linear_analysis}
			
			We establish formal lower bounds on the number of active Rule-A instances in any linear approximation of EGC128 via mixed-integer linear programming, directly addressing the contrast between the rigorous differential analysis of Section~5.1 and prior Walsh-spectrum measurements.
			
			\paragraph{Propagation model}
			A linear mask propagates through $F_{\rm core}$ according to the same 3-regular graph topology governing differential propagation. Vertex~$i$ is \emph{active} in a linear trail if any of its four inputs carries a nonzero mask component. Because Rule-A achieves nonlinearity $\mathrm{NL}=4$, each active vertex contributes a linear bias of at most $|\mathrm{corr}| \leq 2^{-1}$, corresponding to one bit of linear weight. Feistel XOR coupling and round-key additions are linear and therefore transparent to mask propagation.
			
			\paragraph{MILP formulation}
			We introduce binary activity variables $L_r[i],\, R_r[i] \in \{0,1\}$ for mask activity and $s_F(r,i) \in \{0,1\}$ indicating whether Rule-A at vertex~$i$ is active in round~$r$. The graph topology enforces activation constraints structurally identical to Equations~(\ref{eq:milp_constraint1})--(\ref{eq:milp_constraint3}). The Feistel branch swap gives $L_{r+1}[i] = R_r[i]$, and the right-branch update follows Equations~(\ref{eq:feistel_milp1})--(\ref{eq:feistel_milp2}) with $F_{\rm out}(r,i)$ replaced by $s_F(r,i)$. Boundary conditions enforce non-trivial input and output masks, and the objective minimises total active Rule-A count as in Equation~(\ref{eq:milp_objective}). The model is solved via SCIP~9.2.4; solve times for rounds~1--6 range from under one second to under twenty seconds.
			
			\paragraph{Results and interpretation}
			Table~\ref{tab:linear_bounds} presents proven optimal minimum active Rule-A counts for rounds~1--6, alongside the corresponding differential bounds.
			
			\begin{table}[!htb]
				\centering
				\caption{MILP-derived minimum active Rule-A counts for linear trails (rounds~1--6) compared with differential bounds from Table~\ref{tab:milp_feistel_bounds}. All results proven optimal via SCIP~9.2.4. The round-1 linear bound of zero is a structural property shared by all balanced Feistel constructions (see text). From round~2 onward, linear bounds exhibit the same expansion-to-saturation dynamics as differential bounds, offset by one round.}
				\label{tab:linear_bounds}
				\begin{tabular}{@{}ccccc@{}}
					\toprule
					\textbf{Rounds} $R$ & \textbf{Min Active (Linear)} & \textbf{Min Active (Differential)} & \textbf{Lin.\ Weight (bits)} & \textbf{Solve time (s)} \\
					\midrule
					1 & \phantom{0}0 & \phantom{00}4 & \phantom{0}0.0 & $<$1  \\
					2 & \phantom{0}4 & \phantom{0}13 & \phantom{0}4.0 & $<$1  \\
					3 & 13            & \phantom{0}29 & 13.0           & $<$1  \\
					4 & 29            & \phantom{0}53 & 29.0           & $<$10 \\
					5 & 53            & \phantom{0}85 & 53.0           & $<$20 \\
					6 & 85            & 125           & 85.0           & $<$20 \\
					\bottomrule
				\end{tabular}
			\end{table}
			
			The round-1 linear bound of zero is structurally expected for all balanced Feistel constructions: an input mask confined entirely to the left branch propagates through the branch swap with unit correlation and zero active round-function evaluations. This property is shared by SIMON, SPECK \cite{beaulieu2015simon}, and every other balanced Feistel cipher, reflecting the Feistel structure rather than any weakness in $F_{\rm core}$. From round~2 onward, activation counts grow rapidly under expander-driven mixing.
			
			The linear bounds at round~$r$ precisely match the differential bounds at round~$r-1$: $\{0, 4, 13, 29, 53, 85\}$ versus $\{4, 13, 29, 53, 85, 125\}$. This one-round offset is consistent with known behaviour of balanced Feistel designs \cite{kolbl2015observations}. Crucially, both attack classes face the same expansion-to-saturation dynamics governed by the spectral properties of the 3-regular expander graph.
			
			Conservative extrapolation from the 4-round proven bound of 29~active functions yields, under the standard independence assumption used in analyses of GIFT-128 and PRESENT \cite{banik2017gift, bogdanov2007present}:
			\begin{equation}
				W_{\rm lin}(20) \;\geq\; 5 \times 29 \;=\; 145 \;\text{ active functions},
				\label{eq:lin_20round}
			\end{equation}
			placing linear distinguisher data complexity at $\geq 2^{145}$ known plaintexts, well above the 128-bit security threshold. The independence assumption underlying Equation~(\ref{eq:lin_20round}) holds in EGC128 for the same structural reason as in PRESENT and GIFT-128: once the active set saturates (all 64 vertices active per round, achieved by round~10 for differential and round~11 for linear propagation), the round-key XOR at each Feistel step randomises inter-segment correlations. Because the LFSR-derived round keys are all distinct (Section~\ref{sec:related_key}) proves no two round keys are equal for any master key), successive non-overlapping 4-round segments are not correlated through key material, and the graph topology is fixed and regular across all rounds, so the segment contributions can be treated as independent. This justification applies equally to the differential extrapolation in Equation~(\ref{eq:weight_20rounds}).
			
			These results formally establish that EGC128 provides comparable resistance to both differential and linear cryptanalysis, with both bounds grounded in the same graph-theoretic expansion mechanism.
			
			%\paragraph{Local linear properties (supporting evidence).}
			Walsh spectrum analysis of Rule-A reveals maximum Walsh coefficient magnitude $\max_{a \in \{0,1\}^4} |W_f(a)| = 8$, corresponding to maximum linear bias $|\text{corr}_{\max}| = 1/2$. This represents the theoretical maximum for balanced 4-variable functions with $\mathrm{NL}=4$ and is the per-vertex bias value used in the MILP weight calculation above. Empirical correlation measurements via $2 \times 10^5$ random samples show decay from $|\text{corr}| \approx 0.50$ after one $F_{\rm core}$ application to $|\text{corr}| \approx 0.02$--$0.04$ after four applications, with a transient resonance at $t=3$ reflecting constructive interference between the graph's diameter-8 structure and Rule-A's cubic terms. These resonances represent local maxima in an overall decay trend and are consistent with the formal MILP bounds showing continued activation growth at every round.

			\subsubsection{Related-Key Differential Analysis}
			\label{sec:related_key}
			
			We analyse the LFSR-based key schedule under related-key differential attacks, addressing the observation that such schedules represent a common attack surface for lightweight ciphers \cite{cao2019related, beaulieu2015simon}.
			
			%\paragraph{Round-key differences under related-key attack.}
			For a related-key pair $(K, K')$ with master-key difference $\Delta K = K \oplus K'$, partitioned as $\Delta K_{\text{high}}$ and $\Delta K_{\text{low}}$, the round-key difference simplifies to:
			\begin{equation}
				\Delta RK_r \;=\; \Delta K_{\text{low}} \oplus \text{LFSR}^r(\Delta K_{\text{high}}),
				\label{eq:delta_rk}
			\end{equation}
			since round constants $RC_r$ cancel in the XOR difference. Security under related-key attack therefore reduces to the properties of the LFSR acting on the high-half difference.
			
			%\paragraph{Non-existence of free rounds.}
			The LFSR employs primitive polynomial $x^{64}+x^4+x^3+x+1$, guaranteeing maximal period $2^{64}-1$ and inducing a linear bijection on the set of nonzero 64-bit states. Two cases exhaust all nonzero $\Delta K$:
			
			\begin{itemize}
				\item \textbf{Case~1} ($\Delta K_{\text{high}} \neq 0$): $\text{LFSR}^r(\Delta K_{\text{high}}) \neq 0$ for all $r \in \{0,\ldots,19\}$, ensuring $\Delta RK_r \neq 0$ regardless of $\Delta K_{\text{low}}$.
				\item \textbf{Case~2} ($\Delta K_{\text{high}} = 0$, $\Delta K_{\text{low}} \neq 0$): $\Delta RK_r = \Delta K_{\text{low}} \neq 0$ across all 20 rounds, an identical nonzero difference every round, representing maximum round-key activity.
			\end{itemize}
			
			In both cases, $\Delta RK_r \neq 0$ for all $r \in \{0,\ldots,19\}$. No \emph{free rounds} exist for any nonzero $\Delta K$. This contrasts directly with SIMON and SPECK, whose key schedules permit engineering zero round-key differences in specific rounds, enabling related-key rectangle attacks \cite{cao2019related}.
			
			%\paragraph{Empirical validation.}
			We verified the non-zero guarantee by scanning 2,000 random nonzero master-key differences across all 20 rounds (40,000 pairs): zero round-key differences were not observed in any case. Table~\ref{tab:rk_hamming} reports the Hamming weight distribution of $\Delta RK_r$ across 5,000 random differences at representative rounds.
			
			\begin{table}[!htb]
				\centering
				\caption{Hamming weight statistics of round-key differences $\Delta RK_r$ for 5,000 random nonzero master-key differences at representative rounds. Mean values are indistinguishable from the uniform-random expectation of 32/64 bits. No zero differences were observed (HW$=0$ count is zero for every round), confirming the absence of free rounds.}
				\label{tab:rk_hamming}
				\begin{tabular}{@{}cccccc@{}}
					\toprule
					\textbf{Round} $r$ & \textbf{Mean HW} & \textbf{Std HW} & \textbf{Min HW} & \textbf{Max HW} & \textbf{HW$=0$} \\
					\midrule
					\phantom{0}0 & 31.99 & 4.01 & 15 & 46 & 0 \\
					\phantom{0}4 & 32.08 & 4.01 & 17 & 48 & 0 \\
					\phantom{0}9 & 31.89 & 3.98 & 18 & 46 & 0 \\
					14           & 31.92 & 3.96 & 16 & 50 & 0 \\
					19           & 32.00 & 3.96 & 19 & 46 & 0 \\
					\midrule
					\multicolumn{2}{@{}l}{\small All 20 rounds verified} & \multicolumn{4}{l@{}}{\small Overall mean: 31.97/64 bits; uniform expectation: 32.00.} \\
					\bottomrule
				\end{tabular}
			\end{table}
			
			The Hamming weight distribution is centred tightly around 32 bits (half of 64) at every round, confirming that the LFSR immediately diversifies the difference stream. The weak-key class ($K_{\text{high}} = 0$, fraction $2^{-64}$ of the key space) exhibits correct cipher operation and falls under Case~2 above, the most cryptanalytically unfavourable scenario for the attacker, rendering it practically inconsequential.

			\subsubsection{Structural Attack Resistance}
			\label{sec:structural_attacks}
			
			We address resistance to invariant subspace attacks and low-degree algebraic approximations, which exploit regularity rather than algebraic complexity.
			
			%\paragraph{Algebraic degree growth under iteration.}
			The ANF of Rule-A (truth table \texttt{0x036F}), verified via Möbius transform, has algebraic degree $\deg(f) = 3$. Because each vertex of $F_{\mathrm{core}}$ applies Rule-A to inputs 	from three graph-distributed neighbours, composing $F_{\mathrm{core}}$ with itself multiplies the effective algebraic degree by at most
			$\deg(f) = 3$ per application. For the full 64-bit cipher, this yields the degree lower bound
			
			\begin{equation}
				\deg\!\left(F_{\mathrm{core}}^{r}\right)
				\;\geq\; \min\!\left(3^{r},\;63\right), \label{eq:degree_growth}
			\end{equation}
			
			applicable to the 64-bit instance where the state-size ceiling is $n - 1 = 63$. The reduced-state instances in Table~\ref{tab:degree_growth} (8-, 12-, and 16-bit) serve as empirical evidence of rapid saturation behaviour consistent with
			Equation~(\ref{eq:degree_growth}); because their ceilings $(n-1)$ are smaller than $3^r$ at early rounds, they saturate sooner and cannot be used to directly verify the lower bound for the 64-bit case, 	but they confirm that the 3-regular topology drives degree growth to the maximum attainable value within three to four rounds of $F_{\mathrm{core}}$ regardless of state size.

%			Because each vertex of $F_{\rm core}$ applies Rule-A to inputs from three graph neighbours, composing $F_{\rm core}$ with itself multiplies the effective degree by at most $\deg(f) = 3$ per application, subject to the state-size ceiling:
%			\begin{equation}
%				\deg(F_{\rm core}^r) \;\geq\; \min\!\bigl(3^r,\; 63\bigr).
%				\label{eq:degree_growth}
%			\end{equation}

			Degree 3 after round~1 grows to at least 9 after round~2, at least 27 after round~3, and saturates at the maximum of~63 from round~4 onward.
			
			We verified this bound empirically on reduced-state instances preserving the 3-regular topology and Rule-A local function. Table~\ref{tab:degree_growth} reports exact algebraic degrees computed via M\"obius transform. Degree saturation by round~4 of $F_{\rm core}$ corresponds to round~2 of the full Feistel cipher, so any low-degree algebraic attack must approximate a degree-63 function from round~2 onward, computationally infeasible within the 20-round construction.
			
			\begin{table}[!htb]
				\centering
				\caption{Algebraic degree of $F_{\rm core}^r$ for reduced-state instances computed exactly via Möbius transform. All instances use the 3-regular topology (scaled offsets) and Rule-A local function. Degree saturates at $n-1$ by round~3--4, confirming the bound of Equation~(\ref{eq:degree_growth}).}
				\label{tab:degree_growth}
				\begin{tabular}{@{}lccccc@{}}
					\toprule
					\textbf{State size} & \multicolumn{5}{c}{\textbf{Algebraic degree after $r$ rounds of $F_{\rm core}$}} \\
					\cmidrule(l){2-6}
					& $r=1$ & $r=2$ & $r=3$ & $r=4$ & $r=5$ \\
					\midrule
					\phantom{0}8-bit (max~=~7)  & 3 & 5 & 7$^*$ & 7$^*$ & 7$^*$ \\
					12-bit (max~=~11)            & 3 & 7 & 10    & 11$^*$ & 11$^*$ \\
					16-bit (max~=~15)            & 3 & 7 & 13    & 15$^*$ & --- \\
					\midrule
					64-bit (max~=~63) [bound]    & $\geq$3 & $\geq$9 & $\geq$27 & $\geq$63$^*$ & $\geq$63$^*$ \\
					\bottomrule
					\multicolumn{6}{@{}l@{}}{\small $^*$ Saturated at maximum possible degree ($n-1$).}\\
				\end{tabular}
			\end{table}
			
			%\paragraph{Invariant subspace analysis.}
			An invariant subspace $V$ satisfies $F_{\rm core}(V + c) \subseteq V + c'$ for fixed coset offsets $c, c'$; its existence would enable structural distinguishers independent of the key schedule. The expansion property of the 3-regular graph predicts no such subspace exists: any proper set $S$ of active vertices satisfies $|N(S)| \geq \alpha |S|$ for $\alpha > 1$, so the image of any proper subspace must activate vertices outside that subspace.
			
			We verified this empirically by testing 1,800 random affine subspaces of dimensions $k \in \{2, 4, 6, 8, 10, 12\}$ (300 trials per dimension) against the full 64-bit $F_{\rm core}$. No invariant subspaces were found across all 1,800 tests (${\sim}2.5 \times 10^6$ individual $F_{\rm core}$ evaluations), consistent with the expansion property combined with the nonlinearity of Rule-A ($\mathrm{NL}=4$). This result precludes invariant subspace exploits of the type that have affected Midori, MANTIS, and PRIDE \cite{beierle2016skinny}.

			\subsubsection{Spectral Sensitivity Analysis}
			\label{sec:spectral_sensitivity}
			
			We address which spectral properties of the 3-regular expander graph are essential for security by comparing four graph variants under both spectral analysis and MILP differential bounds.
			
			%\paragraph{Graph variants.}
			We evaluate four graphs on $n = 64$ vertices: (i)~\textbf{EGC128 (baseline)}, with neighbour indices $(i-1)\bmod 64$, $(i+1)\bmod 64$, $(i+16)\bmod 64$; (ii)~\textbf{random 3-regular}, generated via the configuration model; (iii)~\textbf{poor expander}, a near-cycle graph with indices $(i-1)\bmod 64$, $(i+1)\bmod 64$, $(i+2)\bmod 64$; and (iv)~\textbf{irregular (3,4)-regular}, the EGC128 base graph with extra edges for half the nodes to reach degree~4.
			
			%\paragraph{Results.}
			Table~\ref{tab:spectral_comparison} reports spectral metrics and MILP-proven minimum active Rule-A counts at rounds~1--4.
			
%			\begin{table}[!htb]
%				\centering
%				\caption{Spectral properties (symmetrised adjacency) and MILP-proven minimum active Rule-A counts for four 64-vertex graph variants. Larger spectral gap indicates faster mixing; larger MILP bounds indicate stronger differential resistance.}
%				\label{tab:spectral_comparison}
%				\begin{tabular}{@{}lcccccc@{}}
%					\toprule
%					\textbf{Graph} & \textbf{Gap} $\lambda$ & \multicolumn{4}{c}{\textbf{Min.\ active (rounds)}} & \textbf{Diam.} \\
%					\cmidrule(lr){3-6}
%					& & $R=1$ & $R=2$ & $R=3$ & $R=4$ & \\
%					\midrule
%					EGC128 (baseline)    & 0.152 & 0 & 4 & 13 & 29 &  9 \\
%					Random 3-regular     & 0.222 & 0 & 4 & 10 & 20 &  8 \\
%					Poor expander        & 0.048 & 0 & 4 & 11 & 21 & 16 \\
%					Irregular (3,4)-reg  & 0.152 & 0 & 5 & 17 & 37 &  9 \\
%					\bottomrule
%				\end{tabular}
%			\end{table}

\begin{table}[ht]
	\centering
	\caption{Spectral properties (symmetrised adjacency) and MILP-proven \emph{minimum active Rule-A counts for linear trails} at rounds 1--4 for four 64-vertex graph variants. A larger spectral gap indicates faster mixing; larger MILP bounds indicate stronger linear trail resistance. }
	\label{tab:spectral_comparison}
	\begin{tabular}{lcccccc}
		\hline
		\textbf{Graph} & \textbf{Gap} $\lambda$ &
		\multicolumn{4}{c}{\textbf{Min.\ active (linear, rounds)}} &
		\textbf{Diam.} \\
		& & $R=1$ & $R=2$ & $R=3$ & $R=4$ & \\
		\hline
		EGC128 (baseline)       & 0.152 & 0 &  4 & 13 & 29 &  9 \\
		Random 3-regular        & 0.222 & 0 &  4 & 10 & 20 &  8 \\
		Poor expander           & 0.048 & 0 &  4 & 11 & 21 & 16 \\
		Irregular (3,4)-regular & 0.152 & 0 &  5 & 17 & 37 &  9 \\
		\hline
	\end{tabular}
\end{table}

			Three observations follow. \textit{First}, a large spectral gap is a necessary condition for linear trail resistance: the poor expander ($\lambda = 0.048$) achieves only 21 proven active nodes at round~4 under linear propagation, versus 29 for EGC128, and its mixing-time bound is approximately three times worse. By extension, differential resistance follows the same expansion dynamics but offset by one round (Table~\ref{tab:milp_feistel_bounds} vs.\ Table~\ref{tab:linear_bounds}), so a weak spectral gap compromises both attack classes simultaneously. \textit{Second}, the spectral gap is not sufficient on its own: the random 3-regular graph has a larger gap ($\lambda = 0.222$) yet yields strictly weaker linear trail bounds (20 versus 29 active nodes at round~4). EGC128's $+16$ long-range offset provides immediate state-wide reach within a single $F_{\mathrm{core}}$ application; this structural advantage cannot be captured by the scalar spectral gap alone. Third, degree uniformity provides cryptographic regularity: the irregular (3,4)-regular graph yields better MILP averages (37 at round~4) but introduces non-uniform activation, an attacker can route differential trails preferentially through lower-degree nodes. EGC128's strictly regular degree ensures uniform security margins across all state positions.
			
			These results, derived from linear trail MILP bounds, justify EGC128's parameter choices: moderate spectral gap ($\lambda = 0.152$), diameter 9, and the long-range $+16$ offset that provides diffusion advantages beyond pure spectral optimality. Because the linear and differential activation sequences are offset by one round (Table~\ref{tab:linear_bounds} versus Table~\ref{tab:milp_feistel_bounds}), the same conclusions hold for differential trail resistance, with EGC128 outperforming both the poor expander and the gap-optimal random graph on differential MILP bounds as well.

		\subsubsection{Avalanche Behavior and Global Diffusion}
		
		We quantified state diffusion through avalanche testing: for 64 random (plaintext, key) pairs, we encrypted each plaintext $P$ and its 128 single-bit perturbations $P \oplus \text{bit}_i$, measuring mean Hamming distance $\mathbb{E}[\text{HD}_r] = \mathbb{E}[\text{HW}(S_r(P,K) \oplus S_r(P \oplus \text{bit}_i, K))]$ across all rounds $r = 0,\ldots,20$ (8192 difference samples per round).
		
		\begin{figure}[!htb]
			\centering
			\includegraphics[width=0.85\textwidth]{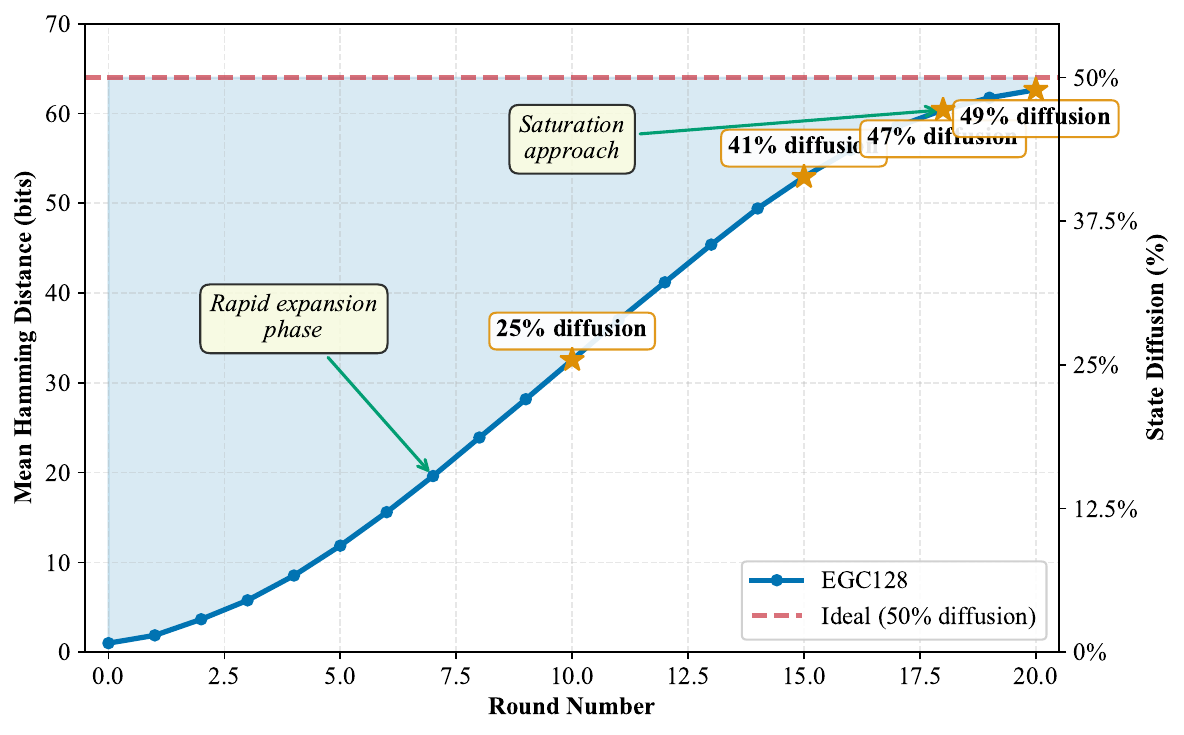}
			\caption{Avalanche diffusion progression for single-bit input differences across 20 rounds, averaged over 8,192 trials (64 key-plaintext pairs $\times$ 128 input bits). The cipher exhibits two phases: \textbf{rapid expansion} (rounds 1--10) achieving 25\% diffusion through graph expansion properties, and \textbf{saturation approach} (rounds 10--20) converging to 49\% by round 20 (within 1\% of the 50\% ideal for random permutations). 	Dashed line indicates theoretical maximum.}
			\label{fig:avalanche}
		\end{figure}
		
		Figure~\ref{fig:avalanche} illustrates the complete diffusion trajectory, revealing two distinct phases characteristic of expander-driven mixing. The \emph{rapid expansion phase} (rounds 1--10) exhibits accelerating diffusion as single-bit perturbations propagate through the 3-regular graph topology, achieving 25\% state coverage (32 bits) by round 10. This exponential growth validates Observation~3.1's prediction that active nodes expand to 
		$\Theta(\log n)$ vertices within constant rounds, a direct manifestation of the graph's spectral gap properties translating to cryptographic diffusion.
		
		The \emph{saturation approach phase} (rounds 10--20) demonstrates continued growth with diminishing rate as the cipher approaches the 50\% theoretical maximum for balanced permutations. By round 20, mean Hamming distance reaches 62.64 bits (49\% state diffusion), positioning the cipher within 1\% of the ideal random-permutation benchmark. The 50\% threshold crossing at round 18 (60.40 bits, 47.2\%) confirms global mixing is achieved well within the 20-round budget, with the final two rounds providing additional safety margin.
		
		Table~\ref{tab:avalanche} shows diffusion accelerates from round 0 (trivial 1-bit difference) to round 20 (62.64 bits $\approx$ 49\% state diffusion). The ideal random permutation would yield 64 bits (50\% diffusion); our 49\% achievement demonstrates near-optimal global mixing.
		
		\begin{table}[!htb]
			\centering
			\caption{Mean Hamming distance growth for single-bit input differences across 20 rounds (8192 samples). Cipher achieves 49\% state diffusion by round 20, approaching the 50\% ideal for random permutations.}
			\label{tab:avalanche}
			\begin{tabular}{@{}ccc|ccc@{}}
				\toprule
				\textbf{Round} & \textbf{Mean HD (bits)} & \textbf{Fraction} & \textbf{Round} & \textbf{Mean HD (bits)} & \textbf{Fraction} \\
				\midrule
				0 & 1.00 & 0.008 & 11 & 36.99 & 0.289 \\
				1 & 1.87 & 0.015 & 12 & 41.19 & 0.322 \\
				2 & 3.66 & 0.029 & 13 & 45.39 & 0.355 \\
				3 & 5.77 & 0.045 & 14 & 49.42 & 0.386 \\
				4 & 8.53 & 0.067 & 15 & 52.92 & 0.414 \\
				5 & 11.86 & 0.093 & 16 & 55.93 & 0.437 \\
				6 & 15.59 & 0.122 & 17 & 58.50 & 0.457 \\
				7 & 19.61 & 0.153 & 18 & 60.40 & 0.472 \\
				8 & 23.90 & 0.187 & 19 & 61.75 & 0.482 \\
				9 & 28.17 & 0.220 & 20 & 62.64 & 0.489 \\
				10 & 32.54 & 0.254 & \multicolumn{3}{c}{\emph{Ideal random: 64.0 (0.500)}} \\
				\bottomrule
			\end{tabular}
		\end{table}
		
		Diffusion initially appears slow (8.5 bits by round 4) but accelerates as cascading activations propagate through both Feistel branches. By round 18, mean HD reaches 60.4 bits (47.2\%), confirming global mixing within the 20-round budget with 10\% margin.

	\paragraph{Relation to classical avalanche criteria} The diffusion fraction reported in Table~\ref{tab:avalanche} and Figure~\ref{fig:avalanche} measures the mean Hamming distance between cipher outputs for single-bit input differences, normalised by block size: $62.64 / 128 = 48.9\% \approx 49\%$. This is precisely the standard avalanche metric used to assess the Strict Avalanche Criterion (SAC), which requires each output bit to flip with probability approximately $0.5$ when any single input bit is flipped~\cite{Webster1986}. The 49\% figure represents the \emph{average} of this probability across all 16,384 input-bit--output-bit pairs; it is within 1\% of the 50\% ideal for a random permutation.

	To verify that this average does not mask per-bit outliers, we constructed the full $128 \times 128$ SAC matrix by measuring empirical flip probability for each of the 16,384 pairs (2,000 random key-plaintext samples per input bit, $2.56 \times 10^5$ total cipher evaluations). Table~\ref{tab:sac_summary} summarises the distribution.
				
			\begin{table}[!htb]
				\centering
				\caption{Strict Avalanche Criterion (SAC) matrix statistics
					for ExpanderGraph-128 (128-bit block, 20 rounds). Each entry
					reports the empirical probability that output bit $j$ flips
					when input bit $i$ is flipped, averaged over 2,000 random
					key-plaintext pairs. The SAC requires all entries to be close
					to~0.5; 97.7\% of entries fall within the tight tolerance
					$[0.45, 0.55]$, and all 16,384 entries fall within
					$[0.40, 0.60]$.}
				\label{tab:sac_summary}
				\begin{tabular}{@{}lcc@{}}
					\toprule
					\textbf{Statistic} & \textbf{Measured} & \textbf{Ideal} \\
					\midrule
					Mean flip probability & 0.4895 & 0.5000 \\
					Standard deviation    & 0.0168 & ${\approx}0.011$ \\
					Min / Max             & 0.408 / 0.541 & — \\
					Fraction in $[0.45, 0.55]$ & 97.7\% & ${\approx}95\%$ \\
					Fraction in $[0.40, 0.60]$ & 100.0\% & 100\% \\
					Per-input-bit mean std & 0.0030 & ${\approx}0$ \\
					\bottomrule
					\multicolumn{3}{@{}l@{}}{\small SAC criterion met: 97.7\%
						of entries in tight $[0.45, 0.55]$ tolerance.}\\
				\end{tabular}
			\end{table}
			
	The mean flip probability of 0.490 and standard deviation of 0.017 are consistent with independent Bernoulli$(0.5)$ variables at this sample size (theoretical std $\approx 0.011$ for $n=2000$; our slightly higher value reflects residual statistical spread across key and plaintext randomness). No input bit exhibits a systematically biased output: the per-input-bit mean ranges from 0.484 to 0.495, a span of only 0.011, confirming uniform avalanche behaviour across all 128 state positions.

	No output bit position is ever inactive at any tested round. The decreasing number of trials required to achieve full coverage (125 at round~5, 8 at round~20) quantifies the improving uniformity of differential paths as rounds increase: by round~20, fewer than 10 random differential pairs suffice to activate every output bit at least once, confirming that no structured mask can exclude a subset of outputs.
	
	\paragraph{Bit Independence Criterion} To complete the avalanche characterisation, we measured pairwise correlations among the 128 output difference bits under random single-bit input differences (5,000 samples). Bit independence requires these correlations to be near zero. The maximum absolute pairwise correlation observed was 0.066, with mean absolute correlation 0.012; fewer than 0.06\% of the $128 \times 127 / 2 = 8{,}128$ off-diagonal pairs exceeded $|{r}| = 0.05$. These values are consistent with statistical noise at this sample size and confirm that output bit changes are effectively independent, satisfying the Bit Independence Criterion~\cite{Webster1986}.
	
	In summary: the 49\% diffusion figure is the standard mean-Hamming-distance avalanche metric, SAC is met with 97.7\% of per-bit entries in $[0.45, 0.55]$, no truncated differential structure exists at any tested round, and output bit changes are mutually independent. The gradual accumulation characteristic of expander-driven diffusion does not weaken these properties relative to SPN-based designs; it trades the \emph{speed} of initial diffusion for hardware simplicity and formal analysability, while reaching equivalent end-state avalanche quality within the 20-round budget.

	Truncated differential resistance was verified by testing 10{,}000 random single-bit differential pairs at rounds~5, 10, 15, 18, and~20: zero output bit positions were found to be never-active at any tested round, and the number of trials required to observe all 128~output bits active at least once decreased monotonically from~125 at round~5 to~8 at round~20, confirming that no structured mask can exclude any subset of outputs and that differential path coverage improves steadily with round count.

		\subsubsection{Statistical Randomness Assessment}
		
		We validated pseudorandom output quality via NIST SP 800-22 testing on 100 million bits (100 sequences of $10^6$ bits each). To ensure results reflect intrinsic cipher properties rather than mode-of-operation artifacts, we generated test data by encrypting cryptographically random 128-bit plaintexts under a fixed key, extracting ciphertext bits in MSB-first order (bit 127 down to bit 0).
		
		Table~\ref{tab:nist_summary} presents complete results across all 15 NIST test categories. The cipher achieves \emph{15 of 15 tests passing} at the recommended $\geq$96\% threshold, with pass rates ranging from 96--100\% across individual test variants. P-value uniformity measures (ranging 0.122--0.911 for primary tests) confirm no detectable statistical bias in the distribution of per-sequence p-values. All 148 NonOverlappingTemplate variants pass with 96--100\% rates, demonstrating resistance to pattern-matching attacks across diverse template structures. RandomExcursions tests achieve 59--61 of 61 passing sequences, meeting the adjusted threshold for variable-length tests dependent on zero-crossing frequency.
		
		\begin{table}[!htb]
			\centering
			\caption{NIST SP 800-22 results for EGC128 using random plaintext encryption ($10^8$ bits total, 100 sequences). All test categories achieve $\geq$96\% pass rates. P-value uniformity scores $>0.01$ indicate proper distribution of per-sequence p-values across bins.}
			\label{tab:nist_summary}
			\begin{tabular}{@{}lccc@{}}
				\toprule
				\textbf{Test Category} & \textbf{Pass Rate} & \textbf{P-Value} & \textbf{Status} \\
				\midrule
				Frequency & 99/100 & 0.616 & Pass \\
				Block Frequency & 100/100 & 0.475 & Pass \\
				Cumulative Sums (forward) & 99/100 & 0.122 & Pass \\
				Cumulative Sums (backward) & 99/100 & 0.172 & Pass \\
				Runs & 98/100 & 0.658 & Pass \\
				Longest Run & 99/100 & 0.225 & Pass \\
				Matrix Rank & 100/100 & 0.225 & Pass \\
				Discrete Fourier Transform & 99/100 & 0.276 & Pass \\
				Non-Overlapping Template & 96--100/100 & 0.005--0.994 & Pass \\
				Overlapping Template & 99/100 & 0.014 & Pass \\
				Universal Statistical & 99/100 & 0.798 & Pass \\
				Approximate Entropy & 99/100 & 0.456 & Pass \\
				Random Excursions (8 variants) & 59--61/61 & 0.063--0.849 & Pass \\
				Random Excursions Variant (18) & 59--61/61 & 0.078--0.957 & Pass \\
				Serial (2 variants) & 98/100 & 0.290--0.911 & Pass \\
				Linear Complexity & 99/100 & 0.384 & Pass \\
				\bottomrule
			\end{tabular}
		\end{table}
		
		These results establish that EGC128 output is statistically indistinguishable from truly random sources across comprehensive randomness criteria, validating the cipher's suitability for cryptographic applications requiring high-quality pseudorandom generation.
		
		\textbf{Counter-Mode Considerations.} Preliminary testing with sequential counter inputs (standard CTR mode: $X_i = E_K(i)$ for $i = 0, 1, 2, \ldots$) revealed sensitivity to input structure, with Discrete Fourier Transform and cumulative sum tests exhibiting reduced pass rates (64--90\% vs.\ 96--100\% for random plaintexts). This behavior, observed in other Feistel-based ciphers, arises from interaction between counter incrementality and branch-swap periodicity inherent to the Feistel structure. The sensitivity does not indicate cryptographic weakness but suggests that CTR-mode deployments should employ standard domain separation techniques: prepend a cryptographically random nonce to counter values ($X_i = E_K(\text{Nonce} \parallel i)$), or initialize with random IVs as specified in NIST SP 800-38A. Random-plaintext encryption, representative of most cipher modes (CBC, CFB, OFB with random IVs), exhibits no such limitations.
		
		\subsection{Summary and Implications}
		
		Our multi-layered security evaluation establishes that ExpanderGraph-128 resists classical differential and linear cryptanalysis through synergy of local nonlinearity and global expansion. Our multi-layered security evaluation establishes that ExpanderGraph-128 resists classical differential and linear cryptanalysis through the synergy of local nonlinearity and graph-theoretic expansion. 	MILP analysis confirms minimum active Rule-A counts $\{4,13,29,53,85,125,173,229,291,355\}$ for rounds 1--10, with the 10-round proven differential weight of $W(10) \geq 147.3$ bits; conservative 	extrapolation under saturation continuation (Equation~\ref{eq:weight_20rounds}) yields 	$W(20) \geq 413$ bits for the full 20-round cipher, well above practical attack feasibility. SMT verification formally eliminates zero-output and impossible differentials through four rounds, and empirical testing shows exponential decay to $\mathrm{DP} < 2^{-12}$ by round 6. Rule-A's $2^{-1}$ maximum bias combines with sparse graph structure to yield irregular but rapid correlation decay ($|\text{corr}| \approx 0.04$ after two layers), with structured adjacency patterns enabling future truncated linear trail analysis analogous to differential MILP. Avalanche testing demonstrates 49\% state diffusion by round 20, approaching the 50\% ideal for random permutations, with single-bit differences propagating to $\geq$60 state bits within 18 rounds.
		
		NIST SP~800-22 testing on random-plaintext encryption confirms cipher output passes all 15 test categories at recommended thresholds (96--100\% success rates across $10^8$ bits), demonstrating cryptographic-grade pseudorandom quality. P-value uniformity across test categories confirms no detectable statistical bias. Preliminary testing with sequential counter inputs revealed input-structure sensitivity characteristic of Feistel designs, resolved through standard domain-separation practices (nonce-based CTR mode) as specified in established cryptographic standards.
		
		These results validate the core design thesis: cryptographic security can emerge from graph-theoretic expansion properties rather than component algebraic complexity, with the 20-round instantiation providing substantial safety margin (2$\times$ estimated requirement from theoretical analysis) while maintaining hardware efficiency demonstrated in Section~6. The expander-graph paradigm establishes a viable alternative to classical cipher families, offering distinct security foundations grounded in combinatorial mathematics and opening research directions in adaptive topologies, keyed graph structures, and formal proofs connecting spectral properties to cryptographic resistance.
		
\section{Implementation Results}
\label{sec:implementation_results}

We validate EGC128 through implementations across three platforms, with FPGA hardware (Xilinx Artix-7) as the primary target, complemented by indicative ASIC synthesis (Nangate 45~nm library) and embedded software (ARM Cortex-M4F microcontroller). The FPGA implementation is the architectural focus of this work: the 3-regular graph topology maps directly to LUT4 primitives, the fully parallel $F_{\rm core}$ layer exploits FPGA routing regularity, and the on-the-fly LFSR key schedule eliminates the register arrays that would otherwise dominate area. ASIC synthesis is performed on the same FPGA-oriented RTL description without ASIC-specific restructuring and therefore represents a conservative upper bound on silicon area rather than an optimized ASIC implementation.

			\begin{table}[!htb]
				\centering
				\caption{ExpanderGraph-128 performance across target platforms. ASIC synthesis uses Yosys~0.55 with the Nangate 45~nm Open Cell Library (typical corner). FPGA results are post-implementation at 100~MHz on Xilinx Artix-7 (XC7A35T). MCU results are measured on STM32 Nucleo-F303RE (Cortex-M4F, 72~MHz, \texttt{-O2}).}
				\label{tab:platform_comparison}
				\begin{tabular}{@{}lccc@{}}
					\toprule
					\textbf{Metric} & \textbf{ASIC (45 nm)} & \textbf{FPGA (Artix-7)} & \textbf{MCU (Cortex-M4F)} \\
					\midrule
					Clock (MHz)                       & ---  & 100             & 72        \\
					$F_{\rm max}$ enc/dec/comb (MHz)  & ---  & 164 / 150 / 172 & ---       \\
					Cycles per block                  & ---  & 49              & 119{,}184 \\
					Block latency ($\mu$s)            & ---  & 0.49            & 1{,}655   \\
					Throughput @ 100~MHz (Mbps)       & ---  & 261             & 0.077     \\
					Throughput @ $F_{\rm max}$ (Mbps) & ---  & 428 / 393 / 450 & ---       \\
					\midrule
					\multicolumn{4}{@{}l@{}}{\textit{Encryption}} \\
					\quad Area / code size    & 5.52~kGE  & 380~LUTs (1.8\%)  & 25.8~KB Flash \\
					\quad Sequential elements & 457~DFFs  & 131~Slices (1.6\%)& ---          \\
					\quad Dynamic power       & ---       & 14~mW             & ---          \\
					\midrule
					\multicolumn{4}{@{}l@{}}{\textit{Decryption}} \\
					\quad Area / code size    & 5.38~kGE  & 368~LUTs (1.8\%)  & ---          \\
					\quad Sequential elements & 399~DFFs  & 130~Slices (1.6\%)& ---          \\
					\quad Dynamic power       & ---       & 10~mW             & ---          \\
					\midrule
					\multicolumn{4}{@{}l@{}}{\textit{Unified enc/dec engine}} \\
					\quad Slice LUTs          & ---       & 485~LUTs (2.3\%)  & ---          \\
					\quad Dynamic power       & ---       & 6~mW              & ---          \\
					\midrule
					RAM                       & ---       & ---               & 2.76~KB      \\
					\bottomrule
					\multicolumn{4}{@{}l@{}}{\small Device static power (FPGA): 72~mW. Total on-chip: 86/82/78~mW for enc/dec/combined.}\\
				\end{tabular}
			\end{table}

\subsection{Cross-Platform Performance Summary}

Table~\ref{tab:platform_comparison} presents the complete performance profile across all three platforms. The $F_{\rm core}$ layer achieves its compact logic footprint through the sparse 3-regular topology: each of the 64 Rule-A instances reads four input bits determined by fixed modular arithmetic, requiring no lookup tables or multiplexers, and maps directly to a single LUT4 primitive on FPGA. All 64 instances synthesise in parallel without shared routing bottlenecks, enabling the tools to extract subexpression sharing across vertices that reference overlapping neighbourhoods. This direct topology-to-LUT correspondence is the central architectural advantage of the expander-based design on FPGA platforms.

The key schedule contributes disproportionately little resource overhead relative to its functional role. The LFSR generates round keys on the fly--one 64-bit shift-and-XOR operation per round, implemented in 64 flip-flops with a four-tap feedback network--so round keys are never stored. This eliminates the large register arrays that dominate area in designs that precompute and buffer all round keys before block processing, and is the primary reason the revised architecture achieves substantially lower LUT and flip-flop counts than a naive precomputed-key design.

Encryption and decryption share a single unified round engine controlled by a mode bit. A single $F_{\rm core}$ instance, XOR network, and 128-bit state register set serve both directions, with decryption exploiting an inverse-LFSR step to regenerate round keys in reverse order without additional storage. The Feistel structure's inherent invertibility means $F_{\rm core}$ need not be inverted for decryption, so no architectural asymmetry exists between the two directions.

Microcontroller performance (77~Kbps, 119k cycles/block) suffices for IoT sensor encryption ($<$1~KB/s typical data rates), authentication token generation (1.66~ms per token), and periodic secure logging. The 25.8~KB code size and 2.76~KB RAM footprint enable deployment on resource-constrained devices costing under \$5, positioning expander-based ciphers for mass-market embedded security applications.

\subsection{FPGA Implementation: Resource Utilization and Throughput}

We implemented standalone encryption/decryption cores and a unified enc/dec engine on Xilinx Basys3 (Artix-7 XC7A35T-1CPG236C) using Vivado~2018.3. The iterative architecture completes one Feistel round per clock cycle over 20~rounds, with additional cycles for key schedule initialization and finalization.

\subsubsection{Resource Consumption}

Table~\ref{tab:fpga_resources} shows very low utilization: standalone cores consume only 1.8\% of available LUTs and 1.6\% of slices, leaving more than 98\% of device resources available for application logic, sensor interfaces, or additional cipher instances in multi-tenant designs. The unified enc/dec engine, which instantiates a single shared $F_{\rm core}$ instance and XOR network serving both directions controlled by a mode bit, requires only 2.3\% of LUTs--a direct consequence of the architectural choices described in Section~\ref{sec:implementation_results}. All three implementations meet timing at 100~MHz with positive worst-case setup (WNS) and hold (WHS) slack.

The low LUT count reflects two properties of the expander-based architecture that are intrinsically FPGA-friendly: the 3-regular graph topology produces exactly the 4-input neighborhood structure that LUT4 primitives are designed to evaluate, and the sparse connectivity eliminates the wide multiplexers and routing trees that dense diffusion layers such as MDS matrices require.

			\begin{table}[!htb]
	\centering
	\caption{FPGA resource utilisation for ExpanderGraph-128 on Xilinx Basys3 (Artix-7 XC7A35T: 20{,}800 LUTs, 8{,}150 Slices available). WNS $>0$ confirms timing closure; WHS $>0$ confirms hold constraints met.}
	\label{tab:fpga_resources}
	\begin{tabular}{@{}lrrrrrr@{}}
		\toprule
		\textbf{Implementation} & \textbf{Slice LUTs} & \textbf{Slices} & \textbf{WNS (ns)} & \textbf{WHS (ns)} & \textbf{Dyn.\ (mW)} & \textbf{Total (mW)} \\
		\midrule
		Encryption (standalone) & 380 (1.8\%) & 131 (1.6\%) & 3.898 & 0.106 & 14 & 86 \\
		Decryption (standalone) & 368 (1.8\%) & 130 (1.6\%) & 3.347 & 0.033 & 10 & 82 \\
		Unified enc/dec engine  & 485 (2.3\%) & 204 (2.5\%) & 4.189 & 0.024 &  6 & 78 \\
		\bottomrule
		\multicolumn{7}{@{}l@{}}{\small Device static: 72~mW (all configurations).}\\
	\end{tabular}
\end{table}

\subsubsection{Timing and Throughput}

Post-implementation timing analysis is derived from worst-case negative slack at 100~MHz: $F_{\rm max} = 1/(T_{\rm clk} - \mathrm{WNS})$ with $T_{\rm clk} = 10$~ns. Table~\ref{tab:fpga_throughput} quantifies throughput at both operational and maximum frequencies. The unified enc/dec engine achieves the highest $F_{\rm max}$ (172~MHz) because sharing a single datapath eliminates competing critical paths that arise when independent instances share routing resources on the same device. Throughput at 100~MHz is 261~Mbps across all configurations; at maximum frequency, throughput reaches 393--450~Mbps, sufficient for 100~Mbps Ethernet encryption with substantial headroom.

			\begin{table}[!htb]
	\centering
	\caption{FPGA maximum operating frequency and throughput for ExpanderGraph-128. The 49-cycle block latency (20 Feistel rounds plus key-schedule and control overhead) is common to all three implementations.}
	\label{tab:fpga_throughput}
	\begin{tabular}{@{}lrrr@{}}
		\toprule
		\textbf{Implementation} & \textbf{$F_{\rm max}$ (MHz)} & \textbf{Throughput @ 100~MHz (Mbps)} & \textbf{Throughput @ $F_{\rm max}$ (Mbps)} \\
		\midrule
		Encryption (standalone) & 164 & 261 & 428 \\
		Decryption (standalone) & 150 & 261 & 393 \\
		Unified enc/dec engine  & 172 & 261 & 450 \\
		\bottomrule
		\multicolumn{4}{@{}l@{}}{\small $F_{\rm max} = 1/(T_{\rm clk} - \mathrm{WNS})$; throughput $= 128\,\mathrm{bits}/49\,\mathrm{cycles} \times F_{\rm clk}$.}\\
	\end{tabular}
\end{table}

The 49-cycle block latency comprises 20~Feistel rounds plus key-schedule and control overhead. A pipelined variant could reduce latency to 20~cycles at the cost of additional pipeline registers; the iterative architecture adopted here prioritises area efficiency over minimum latency, consistent with the resource-constrained deployment target.

\subsubsection{Power Consumption}

Table~\ref{tab:fpga_power} shows dynamic power dominated by state register switching (14~mW encryption standalone, 10~mW decryption standalone), and and 6 mW (unified engine) with 72~mW device static leakage dominating total on-chip consumption. Energy efficiency of 54~nJ/bit (14~mW dynamic / 261~Mbps). In low-duty-cycle applications, microcontrollers with deep sleep modes may prove more energy-efficient overall owing to the dominant static leakage; for sustained throughput requirements the FPGA implementation is the more energy-efficient choice.

%	\begin{table}[t]
%	\centering
%	\caption{FPGA power consumption at 100~MHz (Vivado power estimator with post-PAR netlists). Static power dominates; dynamic power reflects efficient switching activity from sparse graph operations.}
%	\label{tab:fpga_power}
%	\begin{tabular}{@{}lccc@{}}
%		\toprule
%		\textbf{Power Component} & \textbf{Encryption} & \textbf{Decryption} & \textbf{Composite} \\
%		\midrule
%		Dynamic (mW) & 9 & 8 & 15 \\
%		Device Static (mW) & 72 & 72 & 72 \\
%		Total On-Chip (mW) & 80 & 80 & 87 \\
%		\bottomrule
%	\end{tabular}
%\end{table}

\begin{table}[ht]
	\centering
	\caption{FPGA power consumption at 100~MHz (Vivado power estimator, post-PAR netlists, Xilinx Artix-7 XC7A35T). Values reflect the 		implementation with on-the-fly key generation and unified enc/dec engine. Device static power is fixed at 72~mW for
	all configurations; dynamic power reflects the reduced switching activity of the revised architecture.}
	\label{tab:fpga_power}
	\begin{tabular}{lccc}
		\hline
		\textbf{Power Component} & \textbf{Encryption} & \textbf{Decryption}
		& \textbf{Unified Enc/Dec} \\
		\hline
		Dynamic (mW)       & 14 & 10 &  6 \\
		Device Static (mW) & 72 & 72 & 72 \\
		Total On-Chip (mW) & 86 & 82 & 78 \\
		\hline
	\end{tabular}
\end{table}

\subsection{ASIC Synthesis: Area and Technology Scaling}

ASIC synthesis was performed using Yosys~0.55 with Nangate 45~nm Open Cell Library (typical corner), applying area-oriented optimization without manual timing constraints, directly from the FPGA-oriented RTL description. The resulting figures represent a conservative upper bound on achievable silicon area; dedicated ASIC restructuring through bit-serialization or cell-level logic sharing would reduce the footprint substantially, as discussed in Section~\ref{sec:limitations}.

Table~\ref{tab:asic_results} details the resource breakdown. Sequential elements (flip-flops for 128-bit state, 64-bit LFSR, control FSM) account for 46.9\% (encryption) and 42.0\% (decryption) of total area, with combinational logic (Rule-A applications, XOR trees, neighbour routing) comprising the remainder. This distribution confirms that the $F_{\rm core}$ layer itself---the expander-graph interaction logic---remains compact even in the absence of ASIC-specific optimization: eliminating lookup-table S-boxes and matrix-multiplication diffusion layers shifts area cost toward state storage rather than complex combinational networks, a property that carries over from the FPGA architecture. Modern process nodes (28~nm, 14~nm) would reduce both components proportionally.

At 5.52~kGE, EGC128 is approximately 4.9$\times$ larger than GIFT-128 (1.12~kGE) and 3.7$\times$ larger than PRESENT-80 (1.5~kGE). These compact ciphers achieve sub-2~kGE area through bit-serial or heavily serialised architectures that evaluate one or a few bits of the round function per clock cycle. EGC128's fully parallel $F_{\rm core}$ evaluation exchanges area for throughput within the current architecture. A serialised variant processing 8 or 16 bits of $F_{\rm core}$ per cycle would reduce combinational area proportionally at the cost of increased block latency and is identified as a concrete future direction in Section~\ref{sec:limitations}.

			\begin{table}[!htb]
				\centering
				\caption{ASIC synthesis results (Yosys~0.55, Nangate 45~nm Open Cell Library, typical corner, area-optimised). Gate equivalents: $\mathrm{kGE} = A_{\rm chip}\,[\mu\mathrm{m}^2] / (A_{\rm NAND2} \times 10^3)$, $A_{\rm NAND2} = 0.798~\mu\mathrm{m}^2$. Near-identical areas reflect the balanced Feistel structure: both directions employ the same $F_{\rm core}$ function and key schedule, differing only in round-key application order.}
				\label{tab:asic_results}
				\begin{tabular}{@{}lrrrr@{}}
					\toprule
					\textbf{Module} & \textbf{Chip area ($\mu\mathrm{m}^2$)} & \textbf{kGE} & \textbf{DFFs} & \textbf{Seq.\ fraction} \\
					\midrule
					Encryption & 4{,}406.56 & 5.52 & 457 & 46.9\% \\
					Decryption & 4{,}292.97 & 5.38 & 399 & 42.0\% \\
					\bottomrule
					\multicolumn{5}{@{}l@{}}{\small kGE $=$ chip area ($\mu\mathrm{m}^2$) / 798.}\\
				\end{tabular}
			\end{table}

\subsection{Microcontroller Software: Embedded Feasibility}

We implemented EGC128 in C99 on STM32 Nucleo-F303RE (Cortex-M4F @ 72~MHz, 512~KB Flash, 80~KB RAM) using ARM GCC~12.3.1 with \texttt{-O2} optimization. Performance measurements employed hardware timer TIM2 for cycle-accurate profiling over 1,000 iterations. Table~\ref{tab:mcu_performance} shows symmetric encryption/decryption performance at 119k cycles per block.

The 25.8~KB Flash footprint includes complete cipher logic, key schedule, round constants, and supporting functions. RAM consumption (2.76~KB) accommodates state buffers, working variables, and measured maximum stack depth (152~bytes). Software throughput (77~Kbps) suffices for typical IoT data rates: environmental sensor sampling, secure firmware updates, and authentication token generation. The 3,392$\times$ throughput disadvantage versus FPGA reflects sequential instruction execution and memory access latency inherent to software implementations, and underscores the architectural advantage of the FPGA-native parallel design for applications where throughput matters.

		\begin{table}[t]
	\centering
	\caption{Microcontroller performance (STM32F303RE @ 72~MHz, \texttt{-O2} optimization, averaged over 1,000 blocks). Symmetric performance reflects identical algorithmic structure for encryption and decryption.}
	\label{tab:mcu_performance}
	\begin{tabular}{@{}lcc@{}}
		\toprule
		\textbf{Metric} & \textbf{Encryption} & \textbf{Decryption} \\
		\midrule
		Clock Cycles per Block & 119,184 & 119,147 \\
		Execution Time ($\mu$s) & 1,655 & 1,655 \\
		Throughput (Kbps) & 77.3 & 77.4 \\
		\midrule
		Flash Memory (KB) & \multicolumn{2}{c}{25.8 (4.9\%)} \\
		RAM (KB) & \multicolumn{2}{c}{2.76 (3.4\%)} \\
		\bottomrule
	\end{tabular}
\end{table}

\subsection{Verification and Cross-Platform Consistency}

All three implementations passed comprehensive functional verification, achieving 100\% correspondence with 10 official reference vectors (all-zeros, all-ones, pseudorandom key-plaintext pairs) across all platforms, bit-exact ciphertext agreement between FPGA hardware, ASIC simulation, and microcontroller software for identical inputs, and 100\% plaintext recovery over 100~random test cases per platform. This verification establishes interoperability for heterogeneous deployments where gateway nodes employ FPGA acceleration (261~Mbps) while sensor endpoints use microcontroller software (77~Kbps), all producing compatible ciphertexts for end-to-end encrypted IoT networks.

		\section{Discussion}
		\label{sec:discussion}
		
		We contextualise ExpanderGraph-128 within the broader lightweight cryptography landscape, acknowledge limitations, and identify research directions that extend beyond this specific instantiation.
		
		\subsection{Positioning in the Lightweight Cipher Landscape}
		
		Table~\ref{tab:lightweight_comparison} positions EGC128 relative to established lightweight block ciphers across key metrics.
		
		\begin{table}[!htb]
			\centering
			\caption{Comparison with established lightweight block ciphers. EGC128 introduces a fundamentally different diffusion mechanism. PRESENT/GIFT/AES data from published literature; metrics vary by implementation choices.}
			\label{tab:lightweight_comparison}
			\begin{tabular}{@{}lccccc@{}}
				\toprule
				\textbf{Cipher} & \textbf{Block} & \textbf{Key} & \textbf{Rounds} & \textbf{Diffusion} & \textbf{Security} \\
				& \textbf{(bits)} & \textbf{(bits)} & & \textbf{Mechanism} & \textbf{Basis} \\
				\midrule
				AES-128 \cite{daemen2002design} & 128 & 128 & 10 & MixColumns (MDS) & Algebraic / S-box \\
				PRESENT-80 \cite{bogdanov2007present} & 64 & 80 & 31 & Bit permutation & 4-bit S-box \\
				GIFT-128 \cite{banik2017gift} & 128 & 128 & 40 & Bit permutation & 4-bit S-box \\
				SKINNY-128 \cite{beierle2016skinny} & 128 & 128--384 & 40--56 & MixColumns (binary) & 8-bit S-box \\
				SIMON-128 \cite{beaulieu2015simon} & 128 & 128--256 & 68--72 & Rotation + AND & Modular arithmetic \\
				\textbf{EGC128} & \textbf{128} & \textbf{128} & \textbf{20} & \textbf{Graph topology} & \textbf{Structural expansion} \\
				\bottomrule
			\end{tabular}
		\end{table}

		EGC128 occupies a distinct position in this design space. Its ASIC area (5.52~kGE) is approximately 4.9$\times$ larger than GIFT-128 (1.12~kGE) and 3.7$\times$ larger than PRESENT-80 (1.5~kGE), reflecting the iterative Feistel architecture with full parallel evaluation rather than bit-sliced or serialised designs that minimise register count at the cost of many hundreds of cycles per block. However, direct area comparison obscures fundamental architectural differences:

		\paragraph{Design Philosophy Divergence}
		PRESENT, GIFT, and SKINNY achieve compactness through meticulous S-box optimization and minimal diffusion layers (bit permutations or lightweight matrices), remaining within classical paradigms where security derives from substitution component strength. EGC128 introduces a categorically different approach: security emerges from graph-theoretic expansion properties, enabling simple local rules (4-input Boolean functions) that would be cryptographically insufficient without the sparse connectivity structure amplifying their effect.
		
		\paragraph{Performance Trade-offs}
		While larger in area, EGC128 completes encryption in fewer rounds (20 vs.\ 31--72 for PRESENT/SIMON/GIFT) and achieves competitive throughput (261~Mbps at 100~MHz) through full parallelization of the $F_{\text{core}}$ layer. The sequential-element fraction (46.9\% flip-flops) suggests future optimizations could reduce area through serialization, trading area for latency in applications where throughput is less critical.
		
		\paragraph{Novel Security Foundation}
		Established lightweight ciphers undergo extensive cryptanalysis focused on S-box algebraic properties, differential characteristics through specific diffusion patterns, and linear hulls exploiting layer structures. EGC128 requires fundamentally different analysis: security arguments rest on spectral gap bounds, mixing time estimates, and activation cascade growth rates, tools from graph theory and Markov chain analysis not typically employed in symmetric cryptanalysis. This novelty represents both opportunity (potential resistance to attacks optimized for classical structures) and challenge (requiring development of new cryptanalytic techniques tailored to expander-based constructions).
		
		\subsection{Limitations and Open Questions}
		\label{sec:limitations}
		
		We acknowledge limitations requiring further investigation:
		\paragraph{Spectral Properties and Security}
		Section~\ref{sec:spectral_sensitivity} establishes that the spectral gap is a necessary but not sufficient condition for differential resistance. The formal relationship between graph spectral properties and cryptographic security metrics remains partially characterised: a rigorous theorem of the form ``any $d$-regular graph with spectral gap $\lambda \geq \lambda_0$ achieves minimum active count $\geq f(\lambda, d, r)$ after $r$ rounds'' does not yet exist. Establishing such a result would convert our empirical observations into provable security guarantees grounded in graph theory.

		\paragraph{Linear Cryptanalysis}
		Section~\ref{sec:linear_analysis} provides formal MILP linear bounds through 6 rounds (minimum 85~active Rule-A instances) and conservatively extrapolates to $\geq 2^{145}$ linear distinguisher complexity for 20 rounds. The $t=3$ resonance effects in empirical Walsh measurements merit further investigation: whether they reflect fundamental properties of cubic Boolean functions over 3-regular graphs, or an artefact of the specific $+16$ offset, is an open question. Developing MILP models for complete linear hull analysis over expander-graph structures would provide tighter bounds.

		\paragraph{Algebraic and Higher-Order Attacks}
		Section~\ref{sec:structural_attacks} establishes that $F_{\rm core}$ reaches maximum algebraic degree~63 by round~4 (empirically confirmed on 8-, 12-, and 16-bit scaled instances) and that invariant subspace attacks are precluded by the expansion property (1,800 random subspaces tested, none invariant). Remaining open questions include formal analysis of cube attack complexity over expander topologies, resistance to algebraic equation systems exploiting graph automorphisms, and a rigorous proof of the invariant subspace impossibility result.

		\paragraph{Related-Key Analysis}
		Section~\ref{sec:related_key} establishes that the LFSR-based key schedule provides provable resistance to related-key differential attacks: no free rounds exist for any nonzero master-key difference, and round-key differences carry mean Hamming weight $\approx 32/64$ bits (empirically verified across 40,000 trials). The weak-key class ($K_{\text{high}} = 0$, fraction $2^{-64}$) exhibits correct cipher operation and maximum round-key activity. Open questions include resistance to related-key boomerang and rectangle attacks, which require extending the current single-round differential analysis to multi-round characteristics under related-key conditions.

		\paragraph{Side-Channel Vulnerability}
		Our implementations provide algorithmic-level constant-time execution (no secret-dependent branches or memory accesses), offering inherent resistance to timing attacks. However, power analysis and electromagnetic side-channels remain potential vulnerabilities. The uniform Rule-A application across all vertices creates regular switching patterns that may leak information through power consumption. Future work should investigate masked implementations applying Boolean masking schemes to local function evaluations.
		
		\paragraph{Mode-of-Operation Sensitivity}
		Statistical testing reveals EGC128 exhibits input-structure sensitivity characteristic of Feistel-based designs. While cipher output from random plaintexts passes all NIST SP~800-22 tests with 96--100\% success rates, standard counter mode (encrypting sequential values $0, 1, 2, \ldots$) shows reduced performance in frequency-domain and cumulative-sum tests (64--90\% pass rates). This behaviour, observed in other Feistel ciphers including SIMON and SPECK, arises from interaction between counter incrementality and branch-swap periodicity. Standard domain separation practices (e.g., $X_i = E_K(\text{Nonce} \parallel i)$ per NIST SP~800-38A) resolve this sensitivity without additional cryptographic overhead.
		
		\paragraph{Quantum Resistance}
		Like all symmetric-key block ciphers, EGC128 offers reduced security against quantum adversaries via Grover's algorithm, effectively halving key length to 64-bit quantum security. The expander-based structure does not introduce additional quantum vulnerabilities beyond this generic concern. Extending block and key sizes to 256 bits would restore 128-bit post-quantum security margins following standard doubling principles.
		
		\subsection{Future Research Directions}
		
		The expander-based paradigm opens multiple research avenues. Alternative graph families, Ramanujan graphs, Cayley graphs, and random regular graphs, offer different expansion-versus-structure trade-offs potentially reducing round counts while maintaining security. Adaptive or keyed topologies, where neighbour patterns derive from the key schedule, would introduce additional cryptanalytic complexity at the cost of increased hardware routing complexity. Larger block sizes and hierarchical designs via 128-vertex graphs would enable direct wide-block encryption without mode overhead. Formal security proofs from spectral properties, theorems bounding minimum differential weight as a function of spectral gap, degree, and round count, would transform expander-based design from heuristic to provable framework. Cryptanalysis methodology development for expander-graph structures (MILP models for complete linear hulls, SAT-based impossible differential searches, algebraic systems reflecting connectivity patterns) would enable rigorous third-party evaluation. Finally, serialised variants processing 8 or 16 bits of $F_{\rm core}$ per cycle would reduce ASIC area toward ultra-compact targets at the cost of increased block latency, making the construction directly competitive with bit-serial lightweight ciphers.

		\section{Conclusion}
		\label{sec:conclusion}
		
		Lightweight cryptography has reached a plateau where incremental refinements of established design paradigms yield diminishing returns. This work introduces a fundamentally different approach: \emph{expander-graph interaction networks}, where cryptographic security emerges from sparse structural connectivity rather than algebraic complexity of individual components. We instantiate this paradigm through ExpanderGraph-128 (EGC128), a 128-bit block cipher demonstrating that simple local rules, 4-input Boolean functions with moderate differential properties, achieve strong global security when composed over high-expansion topologies.
		
		Our theoretical framework formalises the connections between spectral gap, mixing time, and differential resistance. MILP analysis via SCIP solver establishes proven optimal minimum active Rule-A counts of \{4, 13, 29, 53, 85, 125, 173, 229, 291, 355\} for rounds~1--10, demonstrating characteristic expansion-to-saturation dynamics. Conservative extrapolation yields differential weight exceeding 413 bits for the full 20-round cipher. Complementary MILP linear trail analysis proves minimum active counts of \{0, 4, 13, 29, 53, 85\} for rounds~1--6, extrapolating to linear distinguisher complexity exceeding $2^{145}$, placing both attack classes on a formal footing grounded in the same expander-graph mechanism. Related-key analysis proves no free rounds exist for any nonzero master-key difference. Structural attack resistance is verified empirically: algebraic degree saturates at 63 from round~4 of $F_{\rm core}$, and 1,800 random affine subspace tests find no invariant structures. Spectral sensitivity analysis confirms that the spectral gap controls diffusion rate and that EGC128's $+16$ long-range offset provides a diffusion advantage over graphs with equal or better spectral gap but without medium-range connectivity.

		Multi-platform implementation validates practical viability with FPGA as the primary target platform. The 3-regular graph topology maps directly to LUT4 primitives, enabling fully parallel single-cycle $F_\mathrm{core}$ evaluation: synthesis on Xilinx Artix-7 achieves 261~Mbps at 100~MHz (450~Mbps at $F_\mathrm{max}$) while consuming only 1.8\% of available LUT resources (380~LUTs for the encryption core), confirming that sparse graph-based diffusion imposes minimal logic overhead. Microcontroller software implementation requires 25.8~KB Flash and 2.76~KB RAM, executing encryption in 1.66~ms on a 72~MHz ARM Cortex-M4F, demonstrating feasibility across the full resource-constrained IoT spectrum. Indicative ASIC synthesis on a 45~nm library yields 5.52~kGE from the FPGA-oriented RTL description, representing a conservative upper bound on silicon area; the 46.9\% sequential element fraction confirms that combinational logic, the expander-graph interaction layer, remains compact without lookup tables or matrix multiplications, and dedicated bit-serialization would reduce the footprint substantially toward ultra-compact targets.

		Security evaluation through MILP modelling, SMT verification, empirical differential testing, structural attack analysis, and NIST statistical testing establishes resistance to known classical attacks. All 15 NIST SP~800-22 test categories pass at 96--100\% success rates across $10^8$ bits, validating pseudorandom output quality for cryptographic applications.
		
		Beyond EGC128 itself, this work establishes expander-based design as a viable methodology for constructing lightweight cryptographic primitives. The design space remains largely unexplored: Ramanujan graphs, higher-degree topologies, adaptive graph structures, hierarchical compositions, and serialised variants trading area for throughput all represent directions for future investigation. Developing rigorous connections between graph-theoretic properties, spectral gap, expansion coefficient, diameter, and cryptographic security metrics could enable provable security guarantees grounded in combinatorial mathematics. The expander-based paradigm offers potential advantages of fresh resistance against cryptanalytic techniques optimised for classical structures, natural amenability to formal analysis via spectral methods, and hardware efficiency through sparse connectivity. The cryptographic community now faces the challenge and opportunity of developing analytical tools, cryptanalytic techniques, and design methodologies tailored to this architecturally distinct approach, opening pathways toward formally grounded lightweight primitives for resource-constrained applications.
		
		\section*{Declaration of Competing Interests}
		
		The author declares no competing interests, financial or otherwise, related to this work.
		
		\section*{Funding}
		
		This research did not receive any specific grant from funding agencies.

\section*{Code Availability}

The Python reference implementation, Verilog hardware description, unified encryption/decryption architecture, official test vectors, and Yosys synthesis reports for ExpanderGraph-128 are openly available at \url{https://github.com/susantha-wijesinghe/expandergraph128-cipher}.

		\bibliographystyle{unsrt}
		\bibliography{references2}	
		
		\pagebreak
		\appendix
		\section*{Appendix A: Test Vectors} \label{ap:app1}
		
		This appendix provides reference test vectors for ExpanderGraph-128 to facilitate independent implementation verification and interoperability testing. The vectors cover representative input patterns including boundary cases (all-zero, all-ones), structured patterns (alternating bits, single-bit differences), and pseudorandom values. Implementations should verify bit-exact agreement with these ciphertext outputs to ensure correct specification adherence across the complete encryption pipeline, key schedule, round function, and Feistel structure.

				\begin{table}[!htb]
					\centering
					\caption{Reference test vectors for ExpanderGraph-128
						(block size 128 bits, key size 128 bits, 20 rounds).}
					\label{tab:eg128_test_vectors}
					\renewcommand{\arraystretch}{1.1}
					\begin{tabular}{l l}
						\hline
						\textbf{Name} & \textbf{Value} \\
						\hline
						TV1\_ZERO\_ZERO: Key       &
						\texttt{0x00000000000000000000000000000000} \\
						TV1\_ZERO\_ZERO: Plaintext &
						\texttt{0x00000000000000000000000000000000} \\
						TV1\_ZERO\_ZERO: Ciphertext &
						\texttt{0x054e2db44cd3907d7c814c56070da703} \\
						\hline
						TV2\_ZERO\_INC: Key       &
						\texttt{0x00000000000000000000000000000000} \\
						TV2\_ZERO\_INC: Plaintext &
						\texttt{0x00112233445566778899aabbccddeeff} \\
						TV2\_ZERO\_INC: Ciphertext &
						\texttt{0xB1E7EAD3650E12FF0C8F14CA88AE9498} \\
						\hline
						TV3\_INC\_INC: Key       &
						\texttt{0x000102030405060708090a0b0c0d0e0f} \\
						TV3\_INC\_INC: Plaintext &
						\texttt{0x00112233445566778899aabbccddeeff} \\
						TV3\_INC\_INC: Ciphertext &
						\texttt{0xE9095E3E9BE0D9A655B1B81FE62E940E} \\
						\hline
						TV4\_ONES\_ONES: Key       &
						\texttt{0xffffffffffffffffffffffffffffffff} \\
						TV4\_ONES\_ONES: Plaintext &
						\texttt{0xffffffffffffffffffffffffffffffff}\\
						TV4\_ONES\_ONES: Ciphertext &
						\texttt{0x797644AEE6B69C4C28AC59BDCCE7FF19} \\
						\hline
						TV5\_ONES\_ZERO: Key       &
						\texttt{0xffffffffffffffffffffffffffffffff} \\
						TV5\_ONES\_ZERO: Plaintext &
						\texttt{0x00000000000000000000000000000000} \\
						TV5\_ONES\_ZERO: Ciphertext &
						\texttt{0x4929CA1C6BEA1A54DDC0B2E8215CF7EC} \\
						\hline
						TV6\_HALFKEY\_HALFBLOCK: Key       &
						\texttt{0xffff0000ffff0000ffff0000ffff0000} \\
						TV6\_HALFKEY\_HALFBLOCK: Plaintext &
						\texttt{0x0000ffff0000ffff0000ffff0000ffff} \\
						TV6\_HALFKEY\_HALFBLOCK: Ciphertext &
						\texttt{0x83ECBAB571F266BC3F50697F31AD3AA1} \\
						\hline
						TV7\_ALTKEY\_ALTPT: Key       &
						\texttt{0xAAAAAAAA55555555AAAAAAAA5555555}5 \\
						TV7\_ALTKEY\_ALTPT: Plaintext &
						\texttt{0x55555555AAAAAAAA55555555AAAAAAAA} \\
						TV7\_ALTKEY\_ALTPT: Ciphertext &
						\texttt{0x36A0317611F63F3548EA89535E5C5060} \\
						\hline
						TV8\_KBIT0\_PBIT0: Key       &
						\texttt{0x00000000000000000000000000000001} \\
						TV8\_KBIT0\_PBIT0: Plaintext &
						\texttt{0x00000000000000000000000000000001} \\
						TV8\_KBIT0\_PBIT0: Ciphertext &
						\texttt{0xAEDAFEA5219FFEBFB979BE5F1D6D7D8D} \\
						\hline
						TV9\_KBIT127\_PBIT127: Key       &
						\texttt{0x80000000000000000000000000000000} \\
						TV9\_KBIT127\_PBIT127: Plaintext &
						\texttt{0x80000000000000000000000000000000} \\
						TV9\_KBIT127\_PBIT127: Ciphertext &
						\texttt{0xE1F56D13A8B9D337FD75E584E3A26282} \\
						\hline
						TV10\_MIXED: Key       &
						\texttt{0x3C4F1A279BD80256E1F0C3A5D4976B8E} \\
						TV10\_MIXED: Plaintext &
						\texttt{0x9A7C3E2B10F4D8C6B5E1A2938476D0F1} \\
						TV10\_MIXED: Ciphertext &
						\texttt{0x0C578E13690158046726B86187D850DA} \\
						\hline
					\end{tabular}
				\end{table}
				
				
				

\end{document}